\documentclass[aps,preprintnumbers,floats, prd, twocolumn,longbibliography, nofootinbib]{revtex4-1}
\usepackage{graphicx}
\usepackage{dcolumn}
\usepackage{bm}
\usepackage{amsmath}
\usepackage{amsthm}
\usepackage{multirow}
\usepackage{enumerate}
\usepackage{amsfonts}
\usepackage{ifthen}
\usepackage{psfrag}
\usepackage{slashed}
\usepackage{hyperref}
\usepackage{gensymb}
\usepackage[utf8]{inputenc}
\usepackage{color}
\usepackage{subfigure}
\usepackage{ulem}
\usepackage[utf8]{inputenc}

\newcommand{\be}{\begin{equation}}
\newcommand{\ee}{\end{equation}}
\newcommand{\bea}{\begin{eqnarray}}
\newcommand{\eea}{\end{eqnarray}}

\begin{document} 

\title{Axion and dark photon limits from Crab Nebula high energy gamma-rays}

\author{Xiaojun Bi$^{1,2}$}
\author{Yu Gao$^{1}$}
\email{gaoyu@ihep.ac.cn}
\author{Junguang Guo$^{1,2}$}
\email{guojg@ihep.ac.cn}
\author{Nick Houston$^{3}$}
\author{Tianjun Li$^{4,2}$}
\author{Fangzhou Xu$^{5,4}$}
\author{Xin Zhang$^{6,7}$}

\affiliation{$^1$ Key Laboratory of Particle Astrophysics, Institute of High Energy Physics, Chinese Academy of Sciences, Beijing, 100049, China}
\affiliation{$^2$ School of Physical Sciences, University of Chinese Academy of Sciences, Beijing, 100049, China}
\affiliation{$^3$ Institute of Theoretical Physics, Faculty of Science, Beijing University of Technology, Beijing 100124, China}
\affiliation{$^4$ Key Laboratory of Theoretical Physics, Institute of Theoretical Physics, Chinese Academy of Sciences, Beijing 100190, China}
\affiliation{$^5$ Institute of Modern Physics, Tsinghua University, Beijing 100084, China}
\affiliation{$^6$ Key Laboratory of Computational Astrophysics, National Astronomical Observatories, Chinese Academy of Sciences, Beijing, 100012, China}
\affiliation{$^7$ School of Astronomy and Space Science, University of Chinese Academy of Sciences, Beijing 100049, China}

\begin{abstract}
The observation of cosmic sub-PeV gamma-rays from the Crab Nebula opens up the possibility of testing cosmic ray photon transparency at the multi-hundred TeV scale. 
Assuming no deviation from a source gamma-ray emission due to accelerated electron inverse-Compton scattering, higher event energies can extend constraints on the effects of new physics; we consider oscillation between gamma-rays and axions, plus attenuation effects from gamma-ray absorption in the case of dark photon dark matter. 
Combining the recent AS$\gamma$ and HAWC sub-PeV data with earlier MAGIC and HEGRA data, axion-like particles are most constrained in the $2\times10^{-7} - 6\times10^{-7}$ eV mass range, where the coupling $g_{a\gamma\gamma}$ is constrained to be below ${1.8}\times 10^{-10}$ GeV$^{-1}$. 
Direct scattering from dark photon dark matter limits kinetic mixing $\epsilon\lesssim$ $10^{-3}$ for masses between 0.01 and $1$ eV.
\end{abstract}

\maketitle

\section{Introduction}

Very high energy cosmic photons are crucial astrophysical observational targets, as they can help us to both understand acceleration mechanisms at high energies and identify cosmic ray sources via their directional information. 
However, very high energy gamma rays above 100 TeV are rare occurrences due to both the scarcity of nearby sources, and the attenuation effect of scattering from the cosmic microwave background (CMB) and extragalactic
background light (EBL), reducing their visibility from distant extra-galactic sources~\cite{Franceschini:2017iwq}. 
Recently, the highest sub-PeV gamma rays from the Crab Nebula events were detected by the Tibet AS$\gamma$ experiment~\cite{Amenomori:2019rjd}. 
The Crab Nebula is a well-known high energy gamma ray source, arising possibly due to acceleration processes in the magnetized wind created by the central pulsar. 
High energy gamma rays in the TeV range originating therefrom have been measured by a number of experiments, including HEGRA~\cite{Aharonian:2004gb}, MAGIC~\cite{Aleksic:2014lkm}, HESS~\cite{Abramowski:2013qea}, etc. 
The latest data from Tibet AS$\gamma$~\cite{Amenomori:2019rjd} and HAWC~\cite{Abeysekara:2019edl} increase the observed gamma ray spectrum to energies above 100 TeV.

Besides astrophysical interests, the propagation of extremely high energy gamma rays can test photon interactions from theories beyond the Standard Model (BSM). 
Well-motivated scenarios include photon mixing into low-mass bosonic states like axions~\cite{Raffelt:1987im} and dark photons~\cite{Dienes:1996zr,Abel:2006qt}, photon decay via Lorentz invariance violation~\cite{Coleman:1997xq, AmelinoCamelia:1997gz}, etc. 
The Crab Nebula is a major Galactic source of high energy gamma rays, and the high energy scale of its gamma ray spectrum is utilized for constraining Lorentz invariance violation~\cite{Satunin:2019gsl}.

BSM processes often lead to an energy-dependent photon flux reduction that benefits from higher energy scales of observed gamma rays. 
In this paper we study the attenuation of gamma ray survival probability from two new-physics processes induced by an axion-like pseudoscalar or a dark photon. 
In comparison with previous cosmic ray measurements, we demonstrate that the newly observed high energy gamma ray data enhances the constraints on photon interactions with these new light bosons.

Originally proposed as a natural solution to the strong CP problem~\cite{Preskill:1982cy, Abbott:1982af, Dine:1982ah}, the QCD axion has recently enjoyed increased attention as a non-thermal dark matter candidate~\cite{Hu:2000ke, Hui:2016ltb} within a well-motivated yet evasive parameter space: fulfilling the correct relic density requires the axion mass to be around $10^{-5}$ eV and the decay constant $f\sim 10^{12}$ GeV. 
In addition, generalized axion-like particles (ALPs) are light pseudoscalars that carry a similar $\frac{a}{f}F\tilde{F}$ coupling to photons. 
ALPs are commonly predicted in grand unified and superstring theories but they are not restricted to a particular mass range. 
The mass of the QCD axion is directly related to the scale of its couplings, however for ALPs there is in general no such constraint~\cite{Jaeckel:2010ni}.
Axions and ALPs are being searched for by a number of experiments (see ~\cite{Irastorza:2018dyq} for a recent review). 
For high energy cosmic gamma rays, axions and ALPs can cause oscillation effects through their coupling to two photons in the presence of galactic magnetic fields~\cite{Masso:2006id}, as explored in recent studies on potential spectral distortions from astrophysical gamma ray sources~\cite{Csaki:2003ef, DeAngelis:2007dqd, DeAngelis:2011id, Simet:2007sa, Fairbairn:2009zi, Meyer:2013pny, Dominguez:2011xy, Mirizzi:2009aj, Mirizzi:2007hr, de2009photon, SanchezConde:2009wu, Belikov:2010ma, Abramowski:2013oea, Reesman:2014ova,
Payez:2014xsa, Berenji:2016jji, TheFermi-LAT:2016zue, Meyer:2016wrm, majumdar2017spectral, Galanti:2018myb, Troitsky:2015nxa, Kohri:2017ljt, Liang:2018mqm, Zhang:2018wpc, Libanov:2019fzq, Long:2019nrz, Xia:2019yud, Rubtsov:2014uga, Galanti:2018upl, vogel2017diffuse, Galanti:2018nvl}.

Dark photons ~\cite{Fayet:1980ad, Holdom:1985ag} are the gauge bosons of hidden sector $U(1)$ gauge symmetries under which the Standard Model (SM) particles are not directly charged. 
The dark photons may kinetically mix with the SM photon, allowing normal matter to acquire a small coupling to the mixed state. 
Such a mixing causes photon-dark photon oscillation in the case where the dark photon has nonzero mass, which also enables cosmic photons to scatter off environmental dark photons~\cite{Ruffini:2015oha}, if in particular the dark photon makes up the dark matter in our Universe. 
Both effects attenuate energetic gamma rays over long propagation distances, as explored in a number of studies~\cite{Lobanov:2012pt, Caputo:2020bdy, Mirizzi:2009iz, zechlin2008new}.

In the following Sections~\ref{sect:aA_osc} - \ref{sect:A'_scat} we briefly discuss ALP-photon oscillation and scattering effects by dark photon on gamma rays, respectively. 
In Section~\ref{sect:analysis} we analyze the compilation of Tibet AS$\gamma$, HEGRA, MAGIC and HAWC data, and give new physics limits by testing the attenuation processes, and then finally conclude in Section~\ref{sect:conclusion}.

\section{Photon-ALP oscillation}
\label{sect:aA_osc}

An ALP $a$ couples to photons with the characteristic coupling
\be
{\cal L}_{a\gamma\gamma} = -\frac{1}{4f}a { F}{\tilde{ F}} = \frac{a}{f} \vec{E}\cdot \vec{B},
\label{eq:lagragian}
\ee
where $f$ relates to the axion decay constant $f_a$ by $f^{-1}= c_{\gamma}\alpha/(\pi f_a)$, $\alpha$ is the fine structure constant and $c_{\gamma}$ is a model dependent coefficient dependent on the underlying theory, e.g. $c_{\gamma}= -0.97$ and 0.36 in KSVZ~\cite{Kim:1979if, Shifman:1979if} and DFSZ~\cite{Zhitnitsky:1980tq, Dine:1981rt} models, respectively.
For ALPs, here we focus on $g_{a\gamma\gamma}\equiv f^{-1}$ as the sole effective parameter for phenomenological purposes.
In the presence of an external magnetic field $B$, Eq.~(\ref{eq:lagragian}) becomes a mixing term~\cite{Raffelt:1987im} between the ALP and a photon that allows for oscillation between the ALP and photon polarisations. 
The propagating mode is described by the three component vector $\Psi=(A_1, A_2, a)^{\rm T}$, where $A_{1,2}$ are the photon polarisations in the transverse $\{\hat{x},\hat{y}\}$ plane, and the propagation direction is $\hat{z}$.
The propagation of $\Psi$ is then governed by the equation~\cite{PhysRevD.37.1237}
\be 
( \omega -i\frac {\rm d}{{\rm d} \ell} + {\mathcal M} ) \Psi=0,
\ee
where $\ell$ donates the propagation distance and $\mathcal{M}$ accounts for mixing-induced oscillation and scattering processes. Choosing the transverse projection of the external magnetic field $B_{\rm T}$ to be along the $\hat{y}$ direction, ${\cal M}$ can be written as
\be 
{\cal M} = \left(
\begin{array}{ccc}
    \Delta_{\perp}& & \\
 & \Delta_{\parallel}& \Delta_{a\gamma}\\
 & \Delta_{a\gamma} & \Delta_{\rm aa} \\
\end{array}
\right)
+i\left(
\begin{array}{ccc}
    \Gamma_{\rm BG}& & \\
           &\Gamma_{\rm BG}& \\
           & & 0\\
\end{array}
\right).
\ee
The first term gives photon-ALP mixing, in which
\bea 
\Delta_{\perp}&=&\Delta_{\rm pl}+2\Delta_{\rm QED}+\Delta_{\rm dis}, \nonumber \\
\Delta_{\parallel}&=&\Delta_{\rm pl}+\frac{7}{2}\Delta_{\rm QED}+\Delta_{\rm dis},
\eea and the Faraday effect is neglected. $\Delta_{\rm pl}$ corresponds to an effective photon mass $-\omega_{\rm pl}^2/(2E)$ due to the presence of free charges, where $E$ is the photon energy and $\omega_{\rm pl}=\sqrt{4\pi\alpha n_e/m_e}$ is the plasma frequency with the electron density $n_e$ and electron mass $m_e$.
$\Delta_{\rm QED}=\alpha E/(45\pi) (B/B_{\rm cr})^2$ accounts for the QED vacuum polarisation effect, where the critical magnetic field $B_{\rm cr}$ equals $m^2_e/|e|\sim 4.4\times 10^{13}$ G~\cite{Meyer:2014epa}.
$\Delta_{\rm dis}=44\alpha^2E\rho_{\rm RF}/(135 m^4_e)$ accounts for dispersion effects from photon-photon scattering on environmental radiation field~\cite{Dobrynina:2014qba}, where the energy density $\rho_{\rm RF}$ includes both the CMB and the interstellar radiation field (ISRF) contributions.
The QED vacuum polarisation and dispersion effects evaluate to a small ${\cal O}(10^{-5})$ correction to the survival probability of 100 TeV photons from the Crab Nebula. These effects are practically negligible and we do not consider these effects in the following analysis. ALP parameters include the mixing term $\Delta_{a\gamma}=g_{a\gamma\gamma} B_{\rm T}/2$ and the diagonal term $\Delta_{aa}=-m^2_a/(2E)$, where $m_a$ is the ALP mass.

The second term  $i\Gamma_{\rm BG}$ accounts for the absorption of high energy photons via the $\gamma\gamma\rightarrow e^+e^-$ process, where the absorption rate $\Gamma_{\rm BG}=1/(2\lambda)$ and $\lambda$ is the photon mean free path in the presence of background photons. The energy threshold for $e^+e^-$ production is
\be
E_{\rm th} \sim \frac{2m^2_e}{E_{\rm BG}} \sim 0.5\left(\frac{1{\rm eV}}{E_{\rm BG}}\right){\rm TeV}\,,
\ee
where $E_{\rm BG}$ is the photon energy in the background radiation field, i.e. the CMB and the ISRF. For photons from the Crab Nebula, the ISRF intensity causes a stronger absorption effect and we use the ISRF model in Ref.~\cite{Vernetto:2016alq}. The absorption rate is obtained by the $\gamma\gamma$ scattering cross-section weighted by the background radiation field energy spectrum $n_{\rm BG}$,
\bea 
\lambda^{-1} &=&  \int^{}_{}
    dE_{\rm BG}\frac{dn_{\rm BG}}{dE_{\rm BG}}
    \hat{\sigma}, \\
\hat{\sigma}&=&\int^2_{0} dx\frac{x}{2}\sigma_{\gamma\gamma},
\eea
where $x=1-{\rm cos}\theta_{\gamma\gamma}$, $\theta_{\gamma\gamma}$ is the angle between incident photons, and
the hard scattering $\sigma_{\gamma\gamma}$ is given by
\bea
\sigma_{\gamma\gamma}&=&\frac{3}{16}\sigma_{\rm T}\left(1-\beta^2\right)
\notag\\
&&\qquad \quad\times\left[(3-\beta^4)\ln\frac{1+\beta}{1-\beta}
-2\beta (2-\beta^2)\right], \\
\beta &\equiv &\left(1-4m^2_{e}/s\right)^{1/2},
\eea
where $\sigma_{\rm T}$ is the Thomson cross section and $s=2xEE_{\rm BG}$ is the Mandelstam variable.
For 100 TeV photons from the Crab Nebula, this absorption would result in a loss of $\sim 10^{-2}$ of the photon flux, 
with smaller losses at lower energies.

In case $B_{\rm T}$ is not strictly along $\hat{y}$ but at an angle $\psi$ to $\hat{y}$, the ${\cal M}$ matrix is modified by a rotation,
\bea
\mathcal{M}&=&V(\psi)\mathcal{M}_0V^{\dagger}(\psi), \\
V(\psi)&=&\left(
\begin{array}{ccc}
    {\rm cos}\psi&{\rm sin}\psi &0 \\
    -{\rm sin}\psi&{\rm cos}\psi&0\\
    0  & 0 & 1
\end{array}
\right).
\eea
The Galactic magnetic field consists of a random component with small coherence scales and a large-scale regular component. 
The random component leads to self-cancellation in oscillation and is ignored in our analysis. We consider the regular Galactic $B$-field model in Ref.~\cite{Jansson:2012rt}, and propagate unpolarised photons through a binned distribution of $B_{\rm T}$ between the Crab Nebula and the Earth, where its average magnitude is around 1.3 $\mu $G. For the binned/sliced $B_{\rm T}$ distribution, the magnitude and direction of $B_{\rm T}$ vary between distance-slices, yet within each slice the $B_{\rm T}$ is considered uniform. The photon survival probability is derived by numerically solving the density matrix evolution equation~\cite{Mirizzi:2009aj,DeAngelis:2011id, Galanti:2018nvl}
\begin{eqnarray}
i\frac{{\rm d}\rho}{{\rm d}\ell}=\left[\rho,\mathcal{M}\right]\,,
\end{eqnarray}
where $\rho = \Psi\Psi^\dagger$. For initially unpolarised photons, $\rho(0)$ takes the initial values $1/2~{\rm diag}(1,1,0)$. The final density matrix $\rho(L)$ is the density matrix at Earth where $L$ represents the total distance between Crab Nebula and Earth. After considering the photon-ALP oscillation effect and the absorption effect, the survival probability of a photon $P_{\rm sur.}= \rho_{11}(L)+\rho_{22}(L)$ where $\rho_{11}, \rho_{22}$ represent the first and second diagonal elements in the density matrix.
The observed gamma ray flux is then
\be 
\frac{{\rm d} \phi}{{\rm d} E} = P_{\rm sur.}\cdot \left.\frac{{\rm d} \phi}{{\rm d} E}\right|_{\rm source}.
\label{eq:axion_conversion}
\ee

\section{Photon-dark photon scattering}
\label{sect:A'_scat}

In extensions of the SM featuring dark photons $\gamma'$ with vector $U(1)$ potential $A^\mu_{\rm D}$, their interactions are introduced via terms of the type
\be
	\mathcal{L}_{{\rm SM}\otimes {\rm D}}=-\epsilon e J_\mu^{\rm SM}A_{\rm D}^\mu\,,
\ee
where $\epsilon$ is a dimensionless mixing parameter and $J_\mu^{\rm SM}$ is the SM electromagnetic current \cite{Fortin:2019npr}.
At energies above that of the dark photon mass but below that of the corresponding fermion mass, these operators can be integrated out to yield the familiar low-energy interactions
\be 
{\cal L} \supset -\frac{1}{4}{\cal F}^{\mu\nu} {\cal F}_{\mu\nu} - \frac{\epsilon}{2} {\cal F}^{\mu\nu} {\cal F'}_{\mu\nu} -\frac{1}{4}{\cal F'}^{\mu\nu} {\cal F'}_{\mu\nu}\,.
\ee

\begin{figure}
\includegraphics[scale=1]{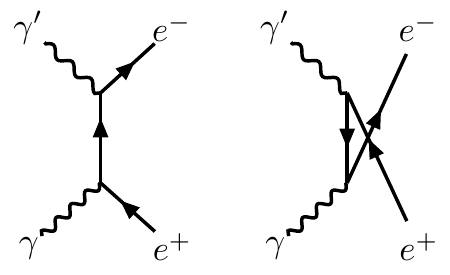}
\caption{Photon attenuation diagrams for $\gamma-\gamma'$ scattering.}
\label{fig:feyn_diag}
\end{figure}

We use $A$ ($A_{\rm D}$) to represent photon (the dark photon) state, respectively. Due to the mixing between dark photon and ordinary photon, we will consider the $A-A_{\rm D}$ scattering case when the dark photon constitutes the dark matter in our Galaxy. 
If the dark photon is massive the scattering process $\gamma \gamma'\rightarrow e^+ e^-$ kinematically opens up for gamma rays above the energy threshold
\be 
E > E'_{\rm th} = \frac{2m_e^2}{m_{\rm D}},
\ee 
and $10^2$ TeV gamma rays reach this threshold for $m_{\rm D}$ down to $10^{-2}$ eV scale.
At the leading order the cosmic ray photon can scatter from the two transversely polarized dark photon modes due to a coupling to the electron via mixing with the QED photon.
The corresponding Feynman diagrams are shown in Fig.~\ref{fig:feyn_diag}. 
The resulting scattering cross-section is
\bea
\sigma_{\rm D} &=& \frac{8\pi\epsilon^2\alpha^2}{3(s-m_{\rm D}^2)^3}\left[-\beta(s^2+4sm_e^2+m_{\rm D}^4)\vphantom{\left(\frac{1+\beta}{1-\beta}\right)}\right.+\ln\left(\frac{1+\beta}{1-\beta}\right)\notag\\
 &&\left.(s^2+4sm_e^2+m_{\rm D}^4-4m_{\rm D}^2m_e^2-8m_e^4)\right], 
\label{eq:cs}
\eea
where $\beta = \sqrt{1-4m_e^2/s}\,$ and $s=2E m_{\rm D}+ m^2_{\rm D}$ is the usual Mandelstam variable.
Eq.~(\ref{eq:cs}) reduces to photon-photon scattering cross-section up to a factor \(2\epsilon^2/3\) when \(m_{\rm D}=0\) due to the absence of the longitudinal mode of the photon. 

If the dark photon makes up the major component of the cold dark matter in our Universe, the corresponding mean free path of photon propagation is 
\be 
\lambda = \frac{1}{n_{\rm D} \sigma_{\rm D}} +(2\Gamma_{\rm BG})^{-1}\,,
\ee
where the dark photon density $n_{\rm D} = \rho_{\rm DM}/m_{\rm D}$ follows from that of the Galactic dark matter distribution, $\rho_{\rm DM}=0.3$ GeV cm$^{-3}$. The second term accounts for absorption due to the background radiation. The observed gamma ray flux is then
\be 
\frac{{\rm d} \phi}{{\rm d} E} = (1-P_{\rm abs.})\cdot \left.\frac{{\rm d} \phi}{{\rm d} E}\right|_{\rm source},
\label{eq:att_scattering}
\ee
where $P_{\rm abs.}$ represents the absorption probability by background photon and dark photon scattering.

\section{Fits to Gamma Rays}
\label{sect:analysis}

\begin{figure}
\includegraphics[scale=0.9]{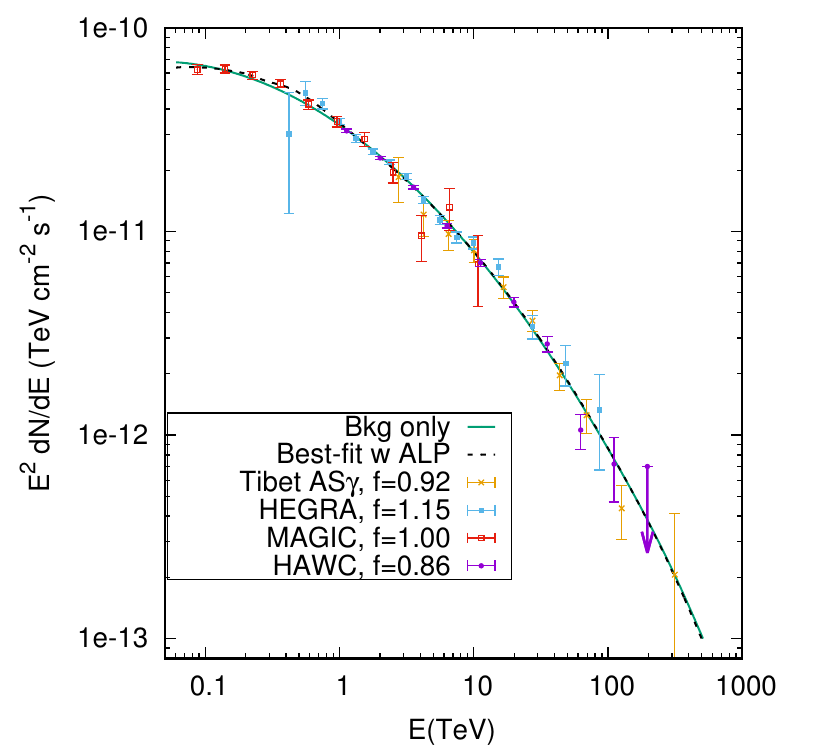}
\caption{
Background-only fit (solid) and the best-fit with ALP effects (dashed) to Tibet AS$\gamma$~\cite{Amenomori:2019rjd}, HAWC~\cite{Abeysekara:2019edl}, HEGRA~\cite{Aharonian:2004gb} and MAGIC~\cite{Aleksic:2014lkm} data. 
The energy scale factor $f$ is allowed to float for each data set and its best-fit value is listed. 
Both the IC background (Eq.~(\ref{eq:parabola})) and ALP-case achieve good fits with $\chi^2=42.2/42$ and $\chi^2=35.6/38$, respectively.
The ALP best-fit is found to be slightly better than background, although they agree to within $1\sigma$.}
\label{fig:bkg_fit}
\end{figure}

To constrain oscillation effects we consider a combination of the recent 100+ TeV gamma ray data from Tibet AS$\gamma$~\cite{Amenomori:2019rjd} and HAWC~\cite{Abeysekara:2019edl}, together with previous measurements from HEGRA~\cite{Aharonian:2004gb} and MAGIC~\cite{Aleksic:2014lkm}. 
The observed gamma ray spectrum is consistent with an expected Inverse-Compton (IC) emission spectrum from accelerated electrons inside the magnetized nebula. 
The shape of such an IC-dominated spectrum is proposed to follow a `parabola' parametrization~\cite{Zaborov:2016jub}
\be 
\frac{{\rm d}\phi^{\rm IC}}{{\rm d}E} = \phi_0 \left( \frac{E}{E_0} \right) ^{\alpha + \beta \log_{10}(E/E_0)},
\label{eq:parabola}
\ee
where the best-fit to AS$\gamma$, HEGRA, MAGIC and HAWC data gives {$\alpha = $-2.57, $\beta = $-0.17}.
This best-fit parameters are obtained by minimizing the $\chi^2$ function of joint fit
\be 
\chi^2 = \sum_{j} \sum_{i} \frac{(\Phi_i^{\rm th} - f^{n-1}_j\cdot\Phi_{j,i})^2}{(f^{n-1}_j\cdot \delta\Phi_{j,i})^2}
+ \sum_{j} \frac{(f_j-1)^2}{(\delta f_j)^2},
\ee
where the subscript $j$ denotes different experimental datasets and $i$ is the $i$th spectral bin in each set. 
$\Phi^{\rm th}$ is the integrated flux $E^n{\rm d}\phi/{\rm d} E$ in each bin (see Appendix~\ref{app:binning} for details), where the raised energy-power index $n$ matches experimental data formats. 
$f_j$ is an energy scale uncertainty that accounts for the significant uncertainty in photon energy reconstruction in air shower measurements, which causes the flux spectrum to effectively shift in energy (and magnitude if $n\neq 1$). 
Such rescaling is typically in the $10\%-20$\% range and it is often necessary for the consistency between experiments. 
The fit with the IC spectrum gives the best scaling factors, as shown in Fig.~\ref{fig:bkg_fit}. 
In our analysis we adopt $\delta f= 0.15$ for HEGRA~\cite{Aharonian:2004gb}, 0.15 for MAGIC, $0.12$ for Tibet AS$\gamma$ and $0.14$ for HAWC. 
We restrict the variation of $f_j$ to be within the range of energy scaling uncertainty $|\Delta f_j|\le \delta f_j$.

\begin{figure}[b!]
\includegraphics[scale=0.85]{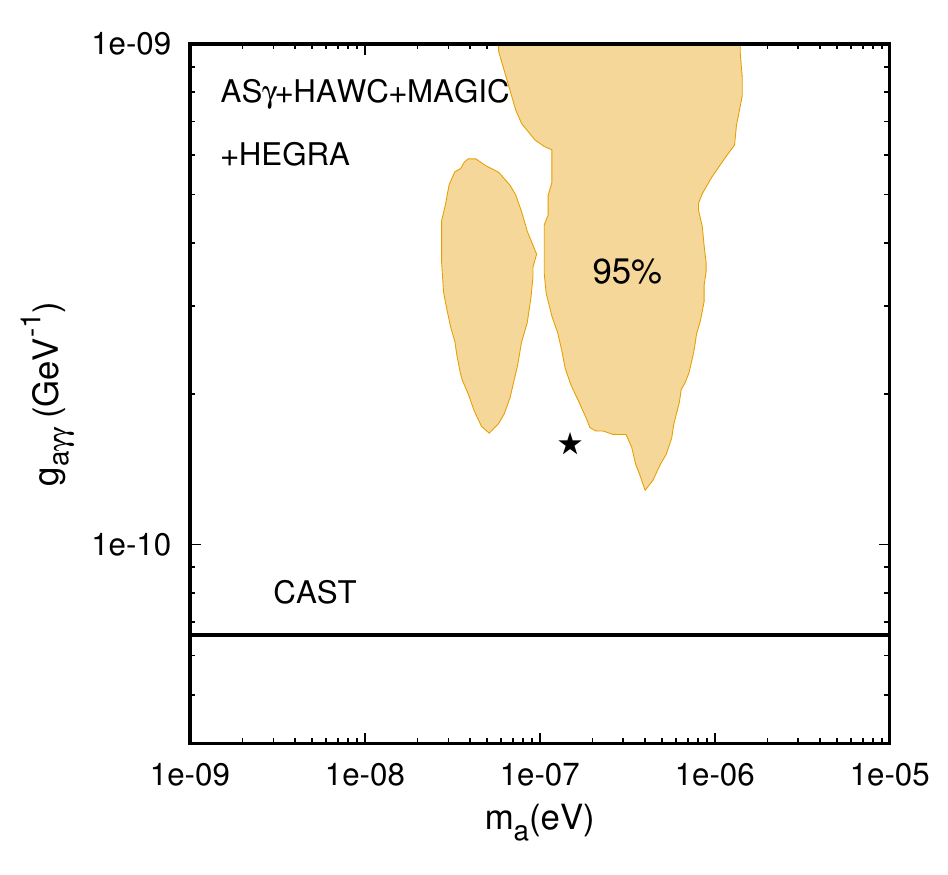}
\caption{
ALP coupling and mass limits from fitting to Tibet AS$\gamma$, HAWC, HEGRA and MAGIC data. The filled contours show 95\% C.L. exclusion regions with the total $\chi^2$= 53.4. The best fit point $\chi^2_{\rm min}$=35.6 is marked by an asterisk. 
The CAST limit~\cite{Anastassopoulos:2017ftl} is shown for comparison.}
\label{fig:fit_axion}
\end{figure}

The fits with axion and dark photon effects are performed after incorporating the photon flux suppression given in Eqs.~(\ref{eq:axion_conversion}) and~(\ref{eq:att_scattering}). 
For ALP-induced oscillation, a best fit point is obtained at $g_{a\gamma\gamma}=1.58\times 10^{-11~}{\rm GeV}^{-1}, m_a=1.26\times 10^{-7~}{\rm eV}$ with the minimal $\chi^2_{\rm min}=35.6$, giving a slight improvement over the background-only fit due to fluctuations in the measured energy spectra.

The fitting process marginalizes over the background IC spectral parameters $\{\phi_0, E_0, \alpha, \beta\}$ and experimental energy scaling factors $\{f_j\}$ to obtain the minimal $\chi^2$ for each point in the ALP $(m_a, g_{a\gamma\gamma})$ and dark photon $(m_{\rm D}, \epsilon )$ parameter spaces. 
The statistical significance of the $\chi^2$ variation with ALP parameters needs special treatment due to the highly oscillatory dependence of the spectral deviation on these parameters. 
Following the statistical prescription for nonlinear dependence in Ref.~\cite{Zhang:2018wpc}, we compare likelihood distribution and find the oscillatory dependence on $g_{a\gamma\gamma}$ and $m_{a}$ is equivalent to 3.5 effective degrees of freedom, which corresponds to a 95\% C.L. increment $\Delta\chi^2\sim 8.7$. 

Interestingly, due to very low global $\chi^2_{\rm min}$ value, even with a $\Delta\chi^2=8.7$ increment, a $\chi^2=44.3$ is only slightly worse (at 78\% C.L.) than $1\sigma$ consistency for a $\chi^2$ distribution with 38 effective degrees of freedom, and this is still a very acceptable global fit. Therefore, we use a more conservative criterion that requires the total $\chi^2$ to be below $53.4$ for 95\% consistency with all the data. The resulting exclusion contours on the $(m_a, g_{a\gamma\gamma})$ plane are shown in Fig.~\ref{fig:fit_axion}.
For 95\% exclusion around an ALP mass $2\times 10^{-7}-6\times10^{-7}$ eV, the gamma ray data give a limit of $g_{a\gamma\gamma}$ below ${1.8}\times 10^{-10}$ GeV$^{-1}$, 
which is close to the latest solar axion constraint from CAST~\cite{Anastassopoulos:2017ftl}.

\begin{figure}
\includegraphics[scale=0.85]{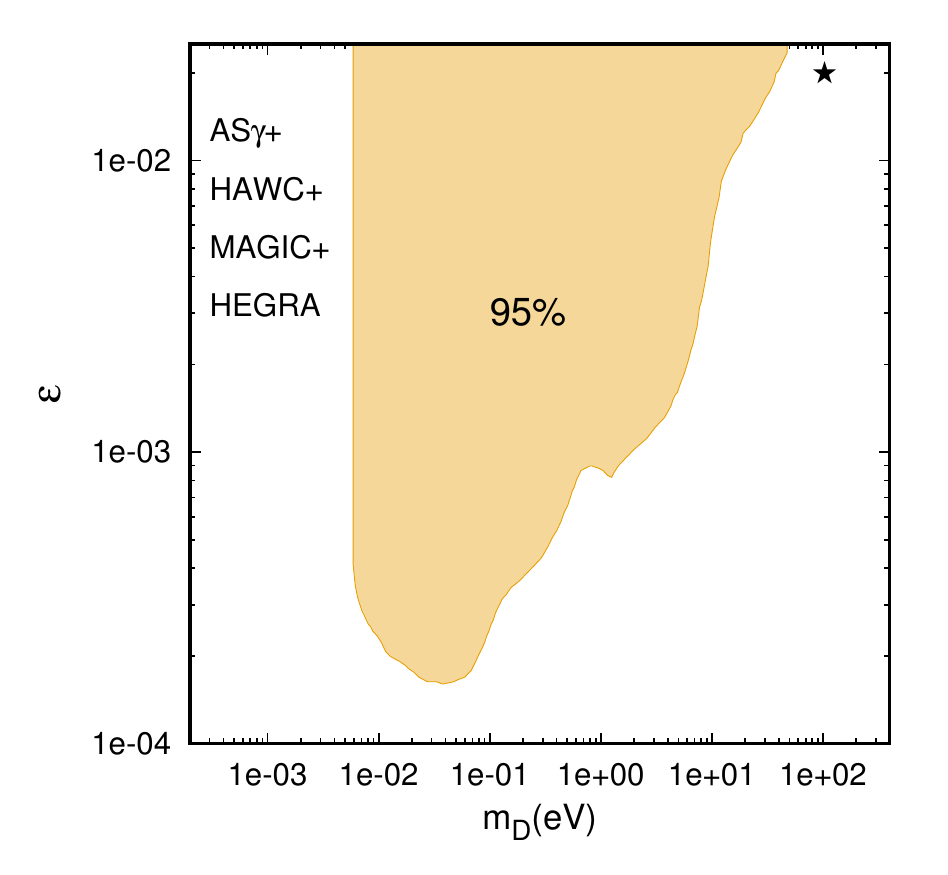}
\caption{
The 95\% C.L. exclusion region for gamma-ray scattering on dark photon as dark matter, from fitting to Tibet AS$\gamma$, HAWC, HEGRA and MAGIC data. The best fit $\chi^2_{\rm min}$=38.0 is marked by the asterisk point.}
\label{fig:fit_dark_photon_sca}
\end{figure}

As the spectral attenuation due to dark photon scattering is non-oscillatory, the 95\% C.L.limit in Fig.~\ref{fig:fit_dark_photon_sca}  corresponds to a total $\chi^2=55.8$. 
The best exclusion of $\epsilon>10^{-3}$ occurs at $10^{-2}~{\rm eV}<m_{\rm D}< 1$ eV. 
These dark photon $\epsilon$ limits are much higher than the typically small mixing required for $A_{\rm D}$ to decouple as dark matter, and significantly less stringent than other dark photon constraints~\cite{essig2013dark}.

Note the HESS experiment also observes high energy gamma rays from the Crab Nebula~\cite{Abramowski:2013qea}, and the resulting spectrum is well fit by the IC background. 
However, the very low $\chi^2$ from HESS data is an overfit to the IC background model, and inclusion of this data set in the combined fit would lead to a less stringent constraint on new physics. Therefore, we do not include HESS in Fig.~\ref{fig:fit_axion}. For comparisons, we list the ALP fitting result to individual data set, and the result after including HESS data, in Appendix~\ref{app:comparisons}.

\section{Conclusion}
\label{sect:conclusion}

We have studied potential gamma ray spectral distortions induced by ALPs and dark photons in light of the recent measurement of photons above 100 TeV from the Tibet AS$\gamma$ and HAWC experiments. 
The newly measured higher energy gamma ray events from the Crab Nebula allow us to extend beyond the sensitivity of previous studies to a higher ALP mass range. 
Assuming an astrophysical background spectrum from accelerated electron inverse Compton occurring scattering inside the Crab Nebula, we performed analyses on the data consistency with the IC background, including ALP-photon oscillation and attenuation effects due to photon-dark photon scattering. 

The Tibet AS$\gamma$, HAWC data and previous HESS, MAGIC and HEGRA data are in very good consistency with a single parabola IC background, at the cost of shifting the energy scale of each experiment in a range comparable to their reported energy uncertainties. 
All of Tibet AS$\gamma$'s highest energy gamma ray events are from the Crab Nebula. 
The relatively close distance to the Earth makes the oscillation effects less significant compared to signals from farther away sources, but the higher energy photons involved can probe into the higher ALP mass range of $10^{-7}-10^{-6}$ eV. 
For a mass region centred at $2\times 10^{-7}-6\times 10^{-7}$ eV, the ALP-photon effective coupling is excluded to $1.8\times 10^{-10}$ GeV$^{-1}$. 
These limits may improve with future accumulation of extremely high energy gamma ray data, or 100+ TeV measurements from other identifiable sources at significant distances. 

We also studied the flux attenuation from scattering on massive dark photons constituting all the dark matter in our Universe, 	
which leads to $\epsilon\lesssim 10^{-3}$ for $10^{-2}~{\rm eV}<m_{\rm D}< 1$ eV. 
This is subdominant to the existing laboratory and cosmological bounds.

\medskip

{\bf Acknowledgments}\\
Y.G. thanks the Institute of High Energy Physics, CAS, for support by the grant no.~Y95461A0U2 and partially by no.~Y7515560U1. T.Li is supported by the National Natural Science Foundation of China under grant no. 11875062 and 11947302, and by the Key Research Program of Frontier Science, CAS. X.J.Bi is supported by NSFC under grant nos.U1738209 and 11851303.

\bibliography{refs}

\begin{thebibliography}{71}%
\makeatletter
\providecommand \@ifxundefined [1]{%
 \@ifx{#1\undefined}
}%
\providecommand \@ifnum [1]{%
 \ifnum #1\expandafter \@firstoftwo
 \else \expandafter \@secondoftwo
 \fi
}%
\providecommand \@ifx [1]{%
 \ifx #1\expandafter \@firstoftwo
 \else \expandafter \@secondoftwo
 \fi
}%
\providecommand \natexlab [1]{#1}%
\providecommand \enquote  [1]{``#1''}%
\providecommand \bibnamefont  [1]{#1}%
\providecommand \bibfnamefont [1]{#1}%
\providecommand \citenamefont [1]{#1}%
\providecommand \href@noop [0]{\@secondoftwo}%
\providecommand \href [0]{\begingroup \@sanitize@url \@href}%
\providecommand \@href[1]{\@@startlink{#1}\@@href}%
\providecommand \@@href[1]{\endgroup#1\@@endlink}%
\providecommand \@sanitize@url [0]{\catcode `\\12\catcode `\$12\catcode
  `\&12\catcode `\#12\catcode `\^12\catcode `\_12\catcode `\%12\relax}%
\providecommand \@@startlink[1]{}%
\providecommand \@@endlink[0]{}%
\providecommand \url  [0]{\begingroup\@sanitize@url \@url }%
\providecommand \@url [1]{\endgroup\@href {#1}{\urlprefix }}%
\providecommand \urlprefix  [0]{URL }%
\providecommand \Eprint [0]{\href }%
\providecommand \doibase [0]{http://dx.doi.org/}%
\providecommand \selectlanguage [0]{\@gobble}%
\providecommand \bibinfo  [0]{\@secondoftwo}%
\providecommand \bibfield  [0]{\@secondoftwo}%
\providecommand \translation [1]{[#1]}%
\providecommand \BibitemOpen [0]{}%
\providecommand \bibitemStop [0]{}%
\providecommand \bibitemNoStop [0]{.\EOS\space}%
\providecommand \EOS [0]{\spacefactor3000\relax}%
\providecommand \BibitemShut  [1]{\csname bibitem#1\endcsname}%
\let\auto@bib@innerbib\@empty
\bibitem [{\citenamefont {Franceschini}\ and\ \citenamefont
  {Rodighiero}(2017)}]{Franceschini:2017iwq}%
  \BibitemOpen
  \bibfield  {author} {\bibinfo {author} {\bibfnamefont {Alberto}\ \bibnamefont
  {Franceschini}}\ and\ \bibinfo {author} {\bibfnamefont {Giulia}\ \bibnamefont
  {Rodighiero}},\ }\bibfield  {title} {\enquote {\bibinfo {title} {{The
  extragalactic background light revisited and the cosmic photon-photon
  opacity}},}\ }\href {\doibase 10.1051/0004-6361/201629684} {\bibfield
  {journal} {\bibinfo  {journal} {Astron. Astrophys.}\ }\textbf {\bibinfo
  {volume} {603}},\ \bibinfo {pages} {A34} (\bibinfo {year} {2017})},\ \Eprint
  {http://arxiv.org/abs/1705.10256} {arXiv:1705.10256 [astro-ph.HE]}
  \BibitemShut {NoStop}%
\bibitem [{\citenamefont {Amenomori}\ \emph {et~al.}(2019)\citenamefont
  {Amenomori} \emph {et~al.}}]{Amenomori:2019rjd}%
  \BibitemOpen
  \bibfield  {author} {\bibinfo {author} {\bibfnamefont {M.}~\bibnamefont
  {Amenomori}} \emph {et~al.},\ }\bibfield  {title} {\enquote {\bibinfo {title}
  {{First Detection of Photons with Energy Beyond 100 TeV from an Astrophysical
  Source}},}\ }\href {\doibase 10.1103/PhysRevLett.123.051101} {\bibfield
  {journal} {\bibinfo  {journal} {Phys. Rev. Lett.}\ }\textbf {\bibinfo
  {volume} {123}},\ \bibinfo {pages} {051101} (\bibinfo {year} {2019})},\
  \Eprint {http://arxiv.org/abs/1906.05521} {arXiv:1906.05521 [astro-ph.HE]}
  \BibitemShut {NoStop}%
\bibitem [{\citenamefont {Aharonian}\ \emph {et~al.}(2004)\citenamefont
  {Aharonian} \emph {et~al.}}]{Aharonian:2004gb}%
  \BibitemOpen
  \bibfield  {author} {\bibinfo {author} {\bibfnamefont {F.}~\bibnamefont
  {Aharonian}} \emph {et~al.} (\bibinfo {collaboration} {HEGRA}),\ }\bibfield
  {title} {\enquote {\bibinfo {title} {{The Crab nebula and pulsar between
  500-GeV and 80-TeV. Observations with the HEGRA stereoscopic air Cerenkov
  telescopes}},}\ }\href {\doibase 10.1086/423931} {\bibfield  {journal}
  {\bibinfo  {journal} {Astrophys. J.}\ }\textbf {\bibinfo {volume} {614}},\
  \bibinfo {pages} {897--913} (\bibinfo {year} {2004})},\ \Eprint
  {http://arxiv.org/abs/astro-ph/0407118} {arXiv:astro-ph/0407118 [astro-ph]}
  \BibitemShut {NoStop}%
\bibitem [{\citenamefont {Aleksić}\ \emph {et~al.}(2016)\citenamefont
  {Aleksić} \emph {et~al.}}]{Aleksic:2014lkm}%
  \BibitemOpen
  \bibfield  {author} {\bibinfo {author} {\bibfnamefont {J.}~\bibnamefont
  {Aleksić}} \emph {et~al.} (\bibinfo {collaboration} {MAGIC}),\ }\bibfield
  {title} {\enquote {\bibinfo {title} {{The major upgrade of the MAGIC
  telescopes, Part II: A performance study using observations of the Crab
  Nebula}},}\ }\href {\doibase 10.1016/j.astropartphys.2015.02.005} {\bibfield
  {journal} {\bibinfo  {journal} {Astropart. Phys.}\ }\textbf {\bibinfo
  {volume} {72}},\ \bibinfo {pages} {76--94} (\bibinfo {year} {2016})},\
  \Eprint {http://arxiv.org/abs/1409.5594} {arXiv:1409.5594 [astro-ph.IM]}
  \BibitemShut {NoStop}%
\bibitem [{\citenamefont {Abramowski}\ \emph {et~al.}(2014)\citenamefont
  {Abramowski} \emph {et~al.}}]{Abramowski:2013qea}%
  \BibitemOpen
  \bibfield  {author} {\bibinfo {author} {\bibfnamefont {A.}~\bibnamefont
  {Abramowski}} \emph {et~al.} (\bibinfo {collaboration} {H.E.S.S.}),\
  }\bibfield  {title} {\enquote {\bibinfo {title} {{H.E.S.S. Observations of
  the Crab during its March 2013 GeV Gamma-Ray Flare}},}\ }\href {\doibase
  10.1051/0004-6361/201323013} {\bibfield  {journal} {\bibinfo  {journal}
  {Astron. Astrophys.}\ }\textbf {\bibinfo {volume} {562}},\ \bibinfo {pages}
  {L4} (\bibinfo {year} {2014})},\ \Eprint {http://arxiv.org/abs/1311.3187}
  {arXiv:1311.3187 [astro-ph.HE]} \BibitemShut {NoStop}%
\bibitem [{\citenamefont {Abeysekara}\ \emph {et~al.}(2019)\citenamefont
  {Abeysekara} \emph {et~al.}}]{Abeysekara:2019edl}%
  \BibitemOpen
  \bibfield  {author} {\bibinfo {author} {\bibfnamefont {A.~U.}\ \bibnamefont
  {Abeysekara}} \emph {et~al.} (\bibinfo {collaboration} {HAWC}),\ }\bibfield
  {title} {\enquote {\bibinfo {title} {{Measurement of the Crab Nebula at the
  Highest Energies with HAWC}},}\ }\href@noop {} {\  (\bibinfo {year}
  {2019})},\ \Eprint {http://arxiv.org/abs/1905.12518} {arXiv:1905.12518
  [astro-ph.HE]} \BibitemShut {NoStop}%
\bibitem [{\citenamefont {Raffelt}\ and\ \citenamefont
  {Stodolsky}(1988{\natexlab{a}})}]{Raffelt:1987im}%
  \BibitemOpen
  \bibfield  {author} {\bibinfo {author} {\bibfnamefont {Georg}\ \bibnamefont
  {Raffelt}}\ and\ \bibinfo {author} {\bibfnamefont {Leo}\ \bibnamefont
  {Stodolsky}},\ }\bibfield  {title} {\enquote {\bibinfo {title} {{Mixing of
  the Photon with Low Mass Particles}},}\ }\href {\doibase
  10.1103/PhysRevD.37.1237} {\bibfield  {journal} {\bibinfo  {journal} {Phys.
  Rev.}\ }\textbf {\bibinfo {volume} {D37}},\ \bibinfo {pages} {1237} (\bibinfo
  {year} {1988}{\natexlab{a}})}\BibitemShut {NoStop}%
\bibitem [{\citenamefont {Dienes}\ \emph {et~al.}(1997)\citenamefont {Dienes},
  \citenamefont {Kolda},\ and\ \citenamefont {March-Russell}}]{Dienes:1996zr}%
  \BibitemOpen
  \bibfield  {author} {\bibinfo {author} {\bibfnamefont {Keith~R.}\
  \bibnamefont {Dienes}}, \bibinfo {author} {\bibfnamefont {Christopher~F.}\
  \bibnamefont {Kolda}}, \ and\ \bibinfo {author} {\bibfnamefont {John}\
  \bibnamefont {March-Russell}},\ }\bibfield  {title} {\enquote {\bibinfo
  {title} {{Kinetic mixing and the supersymmetric gauge hierarchy}},}\ }\href
  {\doibase 10.1016/S0550-3213(97)80028-4, 10.1016/S0550-3213(97)00173-9}
  {\bibfield  {journal} {\bibinfo  {journal} {Nucl. Phys.}\ }\textbf {\bibinfo
  {volume} {B492}},\ \bibinfo {pages} {104--118} (\bibinfo {year} {1997})},\
  \Eprint {http://arxiv.org/abs/hep-ph/9610479} {arXiv:hep-ph/9610479 [hep-ph]}
  \BibitemShut {NoStop}%
\bibitem [{\citenamefont {Abel}\ \emph {et~al.}(2008)\citenamefont {Abel},
  \citenamefont {Jaeckel}, \citenamefont {Khoze},\ and\ \citenamefont
  {Ringwald}}]{Abel:2006qt}%
  \BibitemOpen
  \bibfield  {author} {\bibinfo {author} {\bibfnamefont {Steven~A.}\
  \bibnamefont {Abel}}, \bibinfo {author} {\bibfnamefont {Joerg}\ \bibnamefont
  {Jaeckel}}, \bibinfo {author} {\bibfnamefont {Valentin~V.}\ \bibnamefont
  {Khoze}}, \ and\ \bibinfo {author} {\bibfnamefont {Andreas}\ \bibnamefont
  {Ringwald}},\ }\bibfield  {title} {\enquote {\bibinfo {title} {{Illuminating
  the Hidden Sector of String Theory by Shining Light through a Magnetic
  Field}},}\ }\href {\doibase 10.1016/j.physletb.2008.03.076} {\bibfield
  {journal} {\bibinfo  {journal} {Phys. Lett.}\ }\textbf {\bibinfo {volume}
  {B666}},\ \bibinfo {pages} {66--70} (\bibinfo {year} {2008})},\ \Eprint
  {http://arxiv.org/abs/hep-ph/0608248} {arXiv:hep-ph/0608248 [hep-ph]}
  \BibitemShut {NoStop}%
\bibitem [{\citenamefont {Coleman}\ and\ \citenamefont
  {Glashow}(1997)}]{Coleman:1997xq}%
  \BibitemOpen
  \bibfield  {author} {\bibinfo {author} {\bibfnamefont {Sidney~R.}\
  \bibnamefont {Coleman}}\ and\ \bibinfo {author} {\bibfnamefont {Sheldon~L.}\
  \bibnamefont {Glashow}},\ }\bibfield  {title} {\enquote {\bibinfo {title}
  {{Cosmic ray and neutrino tests of special relativity}},}\ }\href {\doibase
  10.1016/S0370-2693(97)00638-2} {\bibfield  {journal} {\bibinfo  {journal}
  {Phys. Lett.}\ }\textbf {\bibinfo {volume} {B405}},\ \bibinfo {pages}
  {249--252} (\bibinfo {year} {1997})},\ \Eprint
  {http://arxiv.org/abs/hep-ph/9703240} {arXiv:hep-ph/9703240 [hep-ph]}
  \BibitemShut {NoStop}%
\bibitem [{\citenamefont {Amelino-Camelia}\ \emph {et~al.}(1998)\citenamefont
  {Amelino-Camelia}, \citenamefont {Ellis}, \citenamefont {Mavromatos},
  \citenamefont {Nanopoulos},\ and\ \citenamefont
  {Sarkar}}]{AmelinoCamelia:1997gz}%
  \BibitemOpen
  \bibfield  {author} {\bibinfo {author} {\bibfnamefont {G.}~\bibnamefont
  {Amelino-Camelia}}, \bibinfo {author} {\bibfnamefont {John~R.}\ \bibnamefont
  {Ellis}}, \bibinfo {author} {\bibfnamefont {N.~E.}\ \bibnamefont
  {Mavromatos}}, \bibinfo {author} {\bibfnamefont {Dimitri~V.}\ \bibnamefont
  {Nanopoulos}}, \ and\ \bibinfo {author} {\bibfnamefont {Subir}\ \bibnamefont
  {Sarkar}},\ }\bibfield  {title} {\enquote {\bibinfo {title} {{Tests of
  quantum gravity from observations of gamma-ray bursts}},}\ }\href {\doibase
  10.1038/31647} {\bibfield  {journal} {\bibinfo  {journal} {Nature}\ }\textbf
  {\bibinfo {volume} {393}},\ \bibinfo {pages} {763--765} (\bibinfo {year}
  {1998})},\ \Eprint {http://arxiv.org/abs/astro-ph/9712103}
  {arXiv:astro-ph/9712103 [astro-ph]} \BibitemShut {NoStop}%
\bibitem [{\citenamefont {Satunin}(2019)}]{Satunin:2019gsl}%
  \BibitemOpen
  \bibfield  {author} {\bibinfo {author} {\bibfnamefont {Petr}\ \bibnamefont
  {Satunin}},\ }\bibfield  {title} {\enquote {\bibinfo {title} {{New
  constraints on Lorentz Invariance violation from Crab Nebula spectrum beyond
  $100$ TeV}},}\ }\href@noop {} {\  (\bibinfo {year} {2019})},\ \Eprint
  {http://arxiv.org/abs/1906.08221} {arXiv:1906.08221 [astro-ph.HE]}
  \BibitemShut {NoStop}%
\bibitem [{\citenamefont {Preskill}\ \emph {et~al.}(1983)\citenamefont
  {Preskill}, \citenamefont {Wise},\ and\ \citenamefont
  {Wilczek}}]{Preskill:1982cy}%
  \BibitemOpen
  \bibfield  {author} {\bibinfo {author} {\bibfnamefont {John}\ \bibnamefont
  {Preskill}}, \bibinfo {author} {\bibfnamefont {Mark~B.}\ \bibnamefont
  {Wise}}, \ and\ \bibinfo {author} {\bibfnamefont {Frank}\ \bibnamefont
  {Wilczek}},\ }\bibfield  {title} {\enquote {\bibinfo {title} {{Cosmology of
  the Invisible Axion}},}\ }\href {\doibase 10.1016/0370-2693(83)90637-8}
  {\bibfield  {journal} {\bibinfo  {journal} {Phys. Lett.}\ }\textbf {\bibinfo
  {volume} {120B}},\ \bibinfo {pages} {127--132} (\bibinfo {year}
  {1983})}\BibitemShut {NoStop}%
\bibitem [{\citenamefont {Abbott}\ and\ \citenamefont
  {Sikivie}(1983)}]{Abbott:1982af}%
  \BibitemOpen
  \bibfield  {author} {\bibinfo {author} {\bibfnamefont {L.~F.}\ \bibnamefont
  {Abbott}}\ and\ \bibinfo {author} {\bibfnamefont {P.}~\bibnamefont
  {Sikivie}},\ }\bibfield  {title} {\enquote {\bibinfo {title} {{A Cosmological
  Bound on the Invisible Axion}},}\ }\href {\doibase
  10.1016/0370-2693(83)90638-X} {\bibfield  {journal} {\bibinfo  {journal}
  {Phys. Lett.}\ }\textbf {\bibinfo {volume} {120B}},\ \bibinfo {pages}
  {133--136} (\bibinfo {year} {1983})}\BibitemShut {NoStop}%
\bibitem [{\citenamefont {Dine}\ and\ \citenamefont
  {Fischler}(1983)}]{Dine:1982ah}%
  \BibitemOpen
  \bibfield  {author} {\bibinfo {author} {\bibfnamefont {Michael}\ \bibnamefont
  {Dine}}\ and\ \bibinfo {author} {\bibfnamefont {Willy}\ \bibnamefont
  {Fischler}},\ }\bibfield  {title} {\enquote {\bibinfo {title} {{The Not So
  Harmless Axion}},}\ }\href {\doibase 10.1016/0370-2693(83)90639-1} {\bibfield
   {journal} {\bibinfo  {journal} {Phys. Lett.}\ }\textbf {\bibinfo {volume}
  {120B}},\ \bibinfo {pages} {137--141} (\bibinfo {year} {1983})}\BibitemShut
  {NoStop}%
\bibitem [{\citenamefont {Hu}\ \emph {et~al.}(2000)\citenamefont {Hu},
  \citenamefont {Barkana},\ and\ \citenamefont {Gruzinov}}]{Hu:2000ke}%
  \BibitemOpen
  \bibfield  {author} {\bibinfo {author} {\bibfnamefont {Wayne}\ \bibnamefont
  {Hu}}, \bibinfo {author} {\bibfnamefont {Rennan}\ \bibnamefont {Barkana}}, \
  and\ \bibinfo {author} {\bibfnamefont {Andrei}\ \bibnamefont {Gruzinov}},\
  }\bibfield  {title} {\enquote {\bibinfo {title} {{Cold and fuzzy dark
  matter}},}\ }\href {\doibase 10.1103/PhysRevLett.85.1158} {\bibfield
  {journal} {\bibinfo  {journal} {Phys. Rev. Lett.}\ }\textbf {\bibinfo
  {volume} {85}},\ \bibinfo {pages} {1158--1161} (\bibinfo {year} {2000})},\
  \Eprint {http://arxiv.org/abs/astro-ph/0003365} {arXiv:astro-ph/0003365
  [astro-ph]} \BibitemShut {NoStop}%
\bibitem [{\citenamefont {Hui}\ \emph {et~al.}(2017)\citenamefont {Hui},
  \citenamefont {Ostriker}, \citenamefont {Tremaine},\ and\ \citenamefont
  {Witten}}]{Hui:2016ltb}%
  \BibitemOpen
  \bibfield  {author} {\bibinfo {author} {\bibfnamefont {Lam}\ \bibnamefont
  {Hui}}, \bibinfo {author} {\bibfnamefont {Jeremiah~P.}\ \bibnamefont
  {Ostriker}}, \bibinfo {author} {\bibfnamefont {Scott}\ \bibnamefont
  {Tremaine}}, \ and\ \bibinfo {author} {\bibfnamefont {Edward}\ \bibnamefont
  {Witten}},\ }\bibfield  {title} {\enquote {\bibinfo {title} {{Ultralight
  scalars as cosmological dark matter}},}\ }\href {\doibase
  10.1103/PhysRevD.95.043541} {\bibfield  {journal} {\bibinfo  {journal} {Phys.
  Rev.}\ }\textbf {\bibinfo {volume} {D95}},\ \bibinfo {pages} {043541}
  (\bibinfo {year} {2017})},\ \Eprint {http://arxiv.org/abs/1610.08297}
  {arXiv:1610.08297 [astro-ph.CO]} \BibitemShut {NoStop}%
\bibitem [{\citenamefont {Jaeckel}\ and\ \citenamefont
  {Ringwald}(2010)}]{Jaeckel:2010ni}%
  \BibitemOpen
  \bibfield  {author} {\bibinfo {author} {\bibfnamefont {Joerg}\ \bibnamefont
  {Jaeckel}}\ and\ \bibinfo {author} {\bibfnamefont {Andreas}\ \bibnamefont
  {Ringwald}},\ }\bibfield  {title} {\enquote {\bibinfo {title} {{The
  Low-Energy Frontier of Particle Physics}},}\ }\href {\doibase
  10.1146/annurev.nucl.012809.104433} {\bibfield  {journal} {\bibinfo
  {journal} {Ann. Rev. Nucl. Part. Sci.}\ }\textbf {\bibinfo {volume} {60}},\
  \bibinfo {pages} {405--437} (\bibinfo {year} {2010})},\ \Eprint
  {http://arxiv.org/abs/1002.0329} {arXiv:1002.0329 [hep-ph]} \BibitemShut
  {NoStop}%
\bibitem [{\citenamefont {Irastorza}\ and\ \citenamefont
  {Redondo}(2018)}]{Irastorza:2018dyq}%
  \BibitemOpen
  \bibfield  {author} {\bibinfo {author} {\bibfnamefont {Igor~G.}\ \bibnamefont
  {Irastorza}}\ and\ \bibinfo {author} {\bibfnamefont {Javier}\ \bibnamefont
  {Redondo}},\ }\bibfield  {title} {\enquote {\bibinfo {title} {{New
  experimental approaches in the search for axion-like particles}},}\ }\href
  {\doibase 10.1016/j.ppnp.2018.05.003} {\bibfield  {journal} {\bibinfo
  {journal} {Prog. Part. Nucl. Phys.}\ }\textbf {\bibinfo {volume} {102}},\
  \bibinfo {pages} {89--159} (\bibinfo {year} {2018})},\ \Eprint
  {http://arxiv.org/abs/1801.08127} {arXiv:1801.08127 [hep-ph]} \BibitemShut
  {NoStop}%
\bibitem [{\citenamefont {Masso}(2008)}]{Masso:2006id}%
  \BibitemOpen
  \bibfield  {author} {\bibinfo {author} {\bibfnamefont {Eduard}\ \bibnamefont
  {Masso}},\ }\bibfield  {title} {\enquote {\bibinfo {title} {{Axions and their
  relatives}},}\ }\href {\doibase 10.1007/978-3-540-73518-2_5} {\bibfield
  {journal} {\bibinfo  {journal} {Lect. Notes Phys.}\ }\textbf {\bibinfo
  {volume} {741}},\ \bibinfo {pages} {83--94} (\bibinfo {year} {2008})},\
  \Eprint {http://arxiv.org/abs/hep-ph/0607215} {arXiv:hep-ph/0607215}
  \BibitemShut {NoStop}%
\bibitem [{\citenamefont {Csaki}\ \emph {et~al.}(2003)\citenamefont {Csaki},
  \citenamefont {Kaloper}, \citenamefont {Peloso},\ and\ \citenamefont
  {Terning}}]{Csaki:2003ef}%
  \BibitemOpen
  \bibfield  {author} {\bibinfo {author} {\bibfnamefont {Csaba}\ \bibnamefont
  {Csaki}}, \bibinfo {author} {\bibfnamefont {Nemanja}\ \bibnamefont
  {Kaloper}}, \bibinfo {author} {\bibfnamefont {Marco}\ \bibnamefont {Peloso}},
  \ and\ \bibinfo {author} {\bibfnamefont {John}\ \bibnamefont {Terning}},\
  }\bibfield  {title} {\enquote {\bibinfo {title} {{Super GZK photons from
  photon axion mixing}},}\ }\href {\doibase 10.1088/1475-7516/2003/05/005}
  {\bibfield  {journal} {\bibinfo  {journal} {JCAP}\ }\textbf {\bibinfo
  {volume} {0305}},\ \bibinfo {pages} {005} (\bibinfo {year} {2003})},\ \Eprint
  {http://arxiv.org/abs/hep-ph/0302030} {arXiv:hep-ph/0302030 [hep-ph]}
  \BibitemShut {NoStop}%
\bibitem [{\citenamefont {De~Angelis}\ \emph {et~al.}(2007)\citenamefont
  {De~Angelis}, \citenamefont {Roncadelli},\ and\ \citenamefont
  {Mansutti}}]{DeAngelis:2007dqd}%
  \BibitemOpen
  \bibfield  {author} {\bibinfo {author} {\bibfnamefont {Alessandro}\
  \bibnamefont {De~Angelis}}, \bibinfo {author} {\bibfnamefont {Marco}\
  \bibnamefont {Roncadelli}}, \ and\ \bibinfo {author} {\bibfnamefont {Oriana}\
  \bibnamefont {Mansutti}},\ }\bibfield  {title} {\enquote {\bibinfo {title}
  {{Evidence for a new light spin-zero boson from cosmological gamma-ray
  propagation?}}}\ }\href {\doibase 10.1103/PhysRevD.76.121301} {\bibfield
  {journal} {\bibinfo  {journal} {Phys. Rev.}\ }\textbf {\bibinfo {volume}
  {D76}},\ \bibinfo {pages} {121301} (\bibinfo {year} {2007})},\ \Eprint
  {http://arxiv.org/abs/0707.4312} {arXiv:0707.4312 [astro-ph]} \BibitemShut
  {NoStop}%
\bibitem [{\citenamefont {De~Angelis}\ \emph {et~al.}(2011)\citenamefont
  {De~Angelis}, \citenamefont {Galanti},\ and\ \citenamefont
  {Roncadelli}}]{DeAngelis:2011id}%
  \BibitemOpen
  \bibfield  {author} {\bibinfo {author} {\bibfnamefont {Alessandro}\
  \bibnamefont {De~Angelis}}, \bibinfo {author} {\bibfnamefont {Giorgio}\
  \bibnamefont {Galanti}}, \ and\ \bibinfo {author} {\bibfnamefont {Marco}\
  \bibnamefont {Roncadelli}},\ }\bibfield  {title} {\enquote {\bibinfo {title}
  {{Relevance of axion-like particles for very-high-energy astrophysics}},}\
  }\href {\doibase 10.1103/PhysRevD.87.109903, 10.1103/PhysRevD.84.105030}
  {\bibfield  {journal} {\bibinfo  {journal} {Phys. Rev.}\ }\textbf {\bibinfo
  {volume} {D84}},\ \bibinfo {pages} {105030} (\bibinfo {year} {2011})},\
  \bibinfo {note} {[Erratum: Phys. Rev.D87,no.10,109903(2013)]},\ \Eprint
  {http://arxiv.org/abs/1106.1132} {arXiv:1106.1132 [astro-ph.HE]} \BibitemShut
  {NoStop}%
\bibitem [{\citenamefont {Simet}\ \emph {et~al.}(2008)\citenamefont {Simet},
  \citenamefont {Hooper},\ and\ \citenamefont {Serpico}}]{Simet:2007sa}%
  \BibitemOpen
  \bibfield  {author} {\bibinfo {author} {\bibfnamefont {Melanie}\ \bibnamefont
  {Simet}}, \bibinfo {author} {\bibfnamefont {Dan}\ \bibnamefont {Hooper}}, \
  and\ \bibinfo {author} {\bibfnamefont {Pasquale~D.}\ \bibnamefont
  {Serpico}},\ }\bibfield  {title} {\enquote {\bibinfo {title} {{The Milky Way
  as a Kiloparsec-Scale Axionscope}},}\ }\href {\doibase
  10.1103/PhysRevD.77.063001} {\bibfield  {journal} {\bibinfo  {journal} {Phys.
  Rev.}\ }\textbf {\bibinfo {volume} {D77}},\ \bibinfo {pages} {063001}
  (\bibinfo {year} {2008})},\ \Eprint {http://arxiv.org/abs/0712.2825}
  {arXiv:0712.2825 [astro-ph]} \BibitemShut {NoStop}%
\bibitem [{\citenamefont {Fairbairn}\ \emph {et~al.}(2011)\citenamefont
  {Fairbairn}, \citenamefont {Rashba},\ and\ \citenamefont
  {Troitsky}}]{Fairbairn:2009zi}%
  \BibitemOpen
  \bibfield  {author} {\bibinfo {author} {\bibfnamefont {Malcolm}\ \bibnamefont
  {Fairbairn}}, \bibinfo {author} {\bibfnamefont {Timur}\ \bibnamefont
  {Rashba}}, \ and\ \bibinfo {author} {\bibfnamefont {Sergey~V.}\ \bibnamefont
  {Troitsky}},\ }\bibfield  {title} {\enquote {\bibinfo {title} {{Photon-axion
  mixing and ultra-high-energy cosmic rays from BL Lac type objects - Shining
  light through the Universe}},}\ }\href {\doibase 10.1103/PhysRevD.84.125019}
  {\bibfield  {journal} {\bibinfo  {journal} {Phys. Rev.}\ }\textbf {\bibinfo
  {volume} {D84}},\ \bibinfo {pages} {125019} (\bibinfo {year} {2011})},\
  \Eprint {http://arxiv.org/abs/0901.4085} {arXiv:0901.4085 [astro-ph.HE]}
  \BibitemShut {NoStop}%
\bibitem [{\citenamefont {Meyer}\ \emph {et~al.}(2013)\citenamefont {Meyer},
  \citenamefont {Horns},\ and\ \citenamefont {Raue}}]{Meyer:2013pny}%
  \BibitemOpen
  \bibfield  {author} {\bibinfo {author} {\bibfnamefont {Manuel}\ \bibnamefont
  {Meyer}}, \bibinfo {author} {\bibfnamefont {Dieter}\ \bibnamefont {Horns}}, \
  and\ \bibinfo {author} {\bibfnamefont {Martin}\ \bibnamefont {Raue}},\
  }\bibfield  {title} {\enquote {\bibinfo {title} {{First lower limits on the
  photon-axion-like particle coupling from very high energy gamma-ray
  observations}},}\ }\href {\doibase 10.1103/PhysRevD.87.035027} {\bibfield
  {journal} {\bibinfo  {journal} {Phys. Rev.}\ }\textbf {\bibinfo {volume}
  {D87}},\ \bibinfo {pages} {035027} (\bibinfo {year} {2013})},\ \Eprint
  {http://arxiv.org/abs/1302.1208} {arXiv:1302.1208 [astro-ph.HE]} \BibitemShut
  {NoStop}%
\bibitem [{\citenamefont {Dominguez}\ \emph {et~al.}(2011)\citenamefont
  {Dominguez}, \citenamefont {Sanchez-Conde},\ and\ \citenamefont
  {Prada}}]{Dominguez:2011xy}%
  \BibitemOpen
  \bibfield  {author} {\bibinfo {author} {\bibfnamefont {A.}~\bibnamefont
  {Dominguez}}, \bibinfo {author} {\bibfnamefont {M.~A.}\ \bibnamefont
  {Sanchez-Conde}}, \ and\ \bibinfo {author} {\bibfnamefont {F.}~\bibnamefont
  {Prada}},\ }\bibfield  {title} {\enquote {\bibinfo {title} {{Axion-like
  particle imprint in cosmological very-high-energy sources}},}\ }\href
  {\doibase 10.1088/1475-7516/2011/11/020} {\bibfield  {journal} {\bibinfo
  {journal} {JCAP}\ }\textbf {\bibinfo {volume} {1111}},\ \bibinfo {pages}
  {020} (\bibinfo {year} {2011})},\ \Eprint {http://arxiv.org/abs/1106.1860}
  {arXiv:1106.1860 [astro-ph.CO]} \BibitemShut {NoStop}%
\bibitem [{\citenamefont {Mirizzi}\ and\ \citenamefont
  {Montanino}(2009)}]{Mirizzi:2009aj}%
  \BibitemOpen
  \bibfield  {author} {\bibinfo {author} {\bibfnamefont {Alessandro}\
  \bibnamefont {Mirizzi}}\ and\ \bibinfo {author} {\bibfnamefont {Daniele}\
  \bibnamefont {Montanino}},\ }\bibfield  {title} {\enquote {\bibinfo {title}
  {{Stochastic conversions of TeV photons into axion-like particles in
  extragalactic magnetic fields}},}\ }\href {\doibase
  10.1088/1475-7516/2009/12/004} {\bibfield  {journal} {\bibinfo  {journal}
  {JCAP}\ }\textbf {\bibinfo {volume} {0912}},\ \bibinfo {pages} {004}
  (\bibinfo {year} {2009})},\ \Eprint {http://arxiv.org/abs/0911.0015}
  {arXiv:0911.0015 [astro-ph.HE]} \BibitemShut {NoStop}%
\bibitem [{\citenamefont {Mirizzi}\ \emph {et~al.}(2007)\citenamefont
  {Mirizzi}, \citenamefont {Raffelt},\ and\ \citenamefont
  {Serpico}}]{Mirizzi:2007hr}%
  \BibitemOpen
  \bibfield  {author} {\bibinfo {author} {\bibfnamefont {Alessandro}\
  \bibnamefont {Mirizzi}}, \bibinfo {author} {\bibfnamefont {Georg~G.}\
  \bibnamefont {Raffelt}}, \ and\ \bibinfo {author} {\bibfnamefont
  {Pasquale~D.}\ \bibnamefont {Serpico}},\ }\bibfield  {title} {\enquote
  {\bibinfo {title} {{Signatures of Axion-Like Particles in the Spectra of TeV
  Gamma-Ray Sources}},}\ }\href {\doibase 10.1103/PhysRevD.76.023001}
  {\bibfield  {journal} {\bibinfo  {journal} {Phys. Rev.}\ }\textbf {\bibinfo
  {volume} {D76}},\ \bibinfo {pages} {023001} (\bibinfo {year} {2007})},\
  \Eprint {http://arxiv.org/abs/0704.3044} {arXiv:0704.3044 [astro-ph]}
  \BibitemShut {NoStop}%
\bibitem [{\citenamefont {De~Angelis}\ \emph {et~al.}(2009)\citenamefont
  {De~Angelis}, \citenamefont {Mansutti}, \citenamefont {Persic},\ and\
  \citenamefont {Roncadelli}}]{de2009photon}%
  \BibitemOpen
  \bibfield  {author} {\bibinfo {author} {\bibfnamefont {A}~\bibnamefont
  {De~Angelis}}, \bibinfo {author} {\bibfnamefont {O}~\bibnamefont {Mansutti}},
  \bibinfo {author} {\bibfnamefont {M}~\bibnamefont {Persic}}, \ and\ \bibinfo
  {author} {\bibfnamefont {M}~\bibnamefont {Roncadelli}},\ }\bibfield  {title}
  {\enquote {\bibinfo {title} {Photon propagation and the very high energy
  $\gamma$-ray spectra of blazars: how transparent is the universe?}}\
  }\href@noop {} {\bibfield  {journal} {\bibinfo  {journal} {Monthly Notices of
  the Royal Astronomical Society: Letters}\ }\textbf {\bibinfo {volume}
  {394}},\ \bibinfo {pages} {L21--L25} (\bibinfo {year} {2009})}\BibitemShut
  {NoStop}%
\bibitem [{\citenamefont {Sanchez-Conde}\ \emph {et~al.}(2009)\citenamefont
  {Sanchez-Conde}, \citenamefont {Paneque}, \citenamefont {Bloom},
  \citenamefont {Prada},\ and\ \citenamefont
  {Dominguez}}]{SanchezConde:2009wu}%
  \BibitemOpen
  \bibfield  {author} {\bibinfo {author} {\bibfnamefont {M.~A.}\ \bibnamefont
  {Sanchez-Conde}}, \bibinfo {author} {\bibfnamefont {D.}~\bibnamefont
  {Paneque}}, \bibinfo {author} {\bibfnamefont {E.}~\bibnamefont {Bloom}},
  \bibinfo {author} {\bibfnamefont {F.}~\bibnamefont {Prada}}, \ and\ \bibinfo
  {author} {\bibfnamefont {A.}~\bibnamefont {Dominguez}},\ }\bibfield  {title}
  {\enquote {\bibinfo {title} {{Hints of the existence of Axion-Like-Particles
  from the gamma-ray spectra of cosmological sources}},}\ }\href {\doibase
  10.1103/PhysRevD.79.123511} {\bibfield  {journal} {\bibinfo  {journal} {Phys.
  Rev.}\ }\textbf {\bibinfo {volume} {D79}},\ \bibinfo {pages} {123511}
  (\bibinfo {year} {2009})},\ \Eprint {http://arxiv.org/abs/0905.3270}
  {arXiv:0905.3270 [astro-ph.CO]} \BibitemShut {NoStop}%
\bibitem [{\citenamefont {Belikov}\ \emph {et~al.}(2011)\citenamefont
  {Belikov}, \citenamefont {Goodenough},\ and\ \citenamefont
  {Hooper}}]{Belikov:2010ma}%
  \BibitemOpen
  \bibfield  {author} {\bibinfo {author} {\bibfnamefont {Alexander~V.}\
  \bibnamefont {Belikov}}, \bibinfo {author} {\bibfnamefont {Lisa}\
  \bibnamefont {Goodenough}}, \ and\ \bibinfo {author} {\bibfnamefont {Dan}\
  \bibnamefont {Hooper}},\ }\bibfield  {title} {\enquote {\bibinfo {title} {{No
  Indications of Axion-Like Particles From Fermi}},}\ }\href {\doibase
  10.1103/PhysRevD.83.063005} {\bibfield  {journal} {\bibinfo  {journal} {Phys.
  Rev.}\ }\textbf {\bibinfo {volume} {D83}},\ \bibinfo {pages} {063005}
  (\bibinfo {year} {2011})},\ \Eprint {http://arxiv.org/abs/1007.4862}
  {arXiv:1007.4862 [astro-ph.HE]} \BibitemShut {NoStop}%
\bibitem [{\citenamefont {Abramowski}\ \emph {et~al.}(2013)\citenamefont
  {Abramowski} \emph {et~al.}}]{Abramowski:2013oea}%
  \BibitemOpen
  \bibfield  {author} {\bibinfo {author} {\bibfnamefont {A.}~\bibnamefont
  {Abramowski}} \emph {et~al.} (\bibinfo {collaboration} {H.E.S.S.}),\
  }\bibfield  {title} {\enquote {\bibinfo {title} {{Constraints on axionlike
  particles with H.E.S.S. from the irregularity of the PKS 2155-304 energy
  spectrum}},}\ }\href {\doibase 10.1103/PhysRevD.88.102003} {\bibfield
  {journal} {\bibinfo  {journal} {Phys. Rev.}\ }\textbf {\bibinfo {volume}
  {D88}},\ \bibinfo {pages} {102003} (\bibinfo {year} {2013})},\ \Eprint
  {http://arxiv.org/abs/1311.3148} {arXiv:1311.3148 [astro-ph.HE]} \BibitemShut
  {NoStop}%
\bibitem [{\citenamefont {Reesman}\ and\ \citenamefont
  {Walker}(2014)}]{Reesman:2014ova}%
  \BibitemOpen
  \bibfield  {author} {\bibinfo {author} {\bibfnamefont {Rebecca}\ \bibnamefont
  {Reesman}}\ and\ \bibinfo {author} {\bibfnamefont {T.~P.}\ \bibnamefont
  {Walker}},\ }\bibfield  {title} {\enquote {\bibinfo {title} {{Probing the
  Scale of ALP Interactions with Fermi Blazars}},}\ }\href {\doibase
  10.1088/1475-7516/2014/08/021} {\bibfield  {journal} {\bibinfo  {journal}
  {JCAP}\ }\textbf {\bibinfo {volume} {1408}},\ \bibinfo {pages} {021}
  (\bibinfo {year} {2014})},\ \Eprint {http://arxiv.org/abs/1402.2533}
  {arXiv:1402.2533 [astro-ph.HE]} \BibitemShut {NoStop}%
\bibitem [{\citenamefont {Payez}\ \emph {et~al.}(2015)\citenamefont {Payez},
  \citenamefont {Evoli}, \citenamefont {Fischer}, \citenamefont {Giannotti},
  \citenamefont {Mirizzi},\ and\ \citenamefont {Ringwald}}]{Payez:2014xsa}%
  \BibitemOpen
  \bibfield  {author} {\bibinfo {author} {\bibfnamefont {Alexandre}\
  \bibnamefont {Payez}}, \bibinfo {author} {\bibfnamefont {Carmelo}\
  \bibnamefont {Evoli}}, \bibinfo {author} {\bibfnamefont {Tobias}\
  \bibnamefont {Fischer}}, \bibinfo {author} {\bibfnamefont {Maurizio}\
  \bibnamefont {Giannotti}}, \bibinfo {author} {\bibfnamefont {Alessandro}\
  \bibnamefont {Mirizzi}}, \ and\ \bibinfo {author} {\bibfnamefont {Andreas}\
  \bibnamefont {Ringwald}},\ }\bibfield  {title} {\enquote {\bibinfo {title}
  {{Revisiting the SN1987A gamma-ray limit on ultralight axion-like
  particles}},}\ }\href {\doibase 10.1088/1475-7516/2015/02/006} {\bibfield
  {journal} {\bibinfo  {journal} {JCAP}\ }\textbf {\bibinfo {volume} {1502}},\
  \bibinfo {pages} {006} (\bibinfo {year} {2015})},\ \Eprint
  {http://arxiv.org/abs/1410.3747} {arXiv:1410.3747 [astro-ph.HE]} \BibitemShut
  {NoStop}%
\bibitem [{\citenamefont {Berenji}\ \emph {et~al.}(2016)\citenamefont
  {Berenji}, \citenamefont {Gaskins},\ and\ \citenamefont
  {Meyer}}]{Berenji:2016jji}%
  \BibitemOpen
  \bibfield  {author} {\bibinfo {author} {\bibfnamefont {Bijan}\ \bibnamefont
  {Berenji}}, \bibinfo {author} {\bibfnamefont {Jennifer}\ \bibnamefont
  {Gaskins}}, \ and\ \bibinfo {author} {\bibfnamefont {Manuel}\ \bibnamefont
  {Meyer}},\ }\bibfield  {title} {\enquote {\bibinfo {title} {{Constraints on
  Axions and Axionlike Particles from Fermi Large Area Telescope Observations
  of Neutron Stars}},}\ }\href {\doibase 10.1103/PhysRevD.93.045019} {\bibfield
   {journal} {\bibinfo  {journal} {Phys. Rev.}\ }\textbf {\bibinfo {volume}
  {D93}},\ \bibinfo {pages} {045019} (\bibinfo {year} {2016})},\ \Eprint
  {http://arxiv.org/abs/1602.00091} {arXiv:1602.00091 [astro-ph.HE]}
  \BibitemShut {NoStop}%
\bibitem [{\citenamefont {Ajello}\ \emph {et~al.}(2016)\citenamefont {Ajello}
  \emph {et~al.}}]{TheFermi-LAT:2016zue}%
  \BibitemOpen
  \bibfield  {author} {\bibinfo {author} {\bibfnamefont {M.}~\bibnamefont
  {Ajello}} \emph {et~al.} (\bibinfo {collaboration} {Fermi-LAT}),\ }\bibfield
  {title} {\enquote {\bibinfo {title} {{Search for Spectral Irregularities due
  to Photon\textendash{}Axionlike-Particle Oscillations with the Fermi Large
  Area Telescope}},}\ }\href {\doibase 10.1103/PhysRevLett.116.161101}
  {\bibfield  {journal} {\bibinfo  {journal} {Phys. Rev. Lett.}\ }\textbf
  {\bibinfo {volume} {116}},\ \bibinfo {pages} {161101} (\bibinfo {year}
  {2016})},\ \Eprint {http://arxiv.org/abs/1603.06978} {arXiv:1603.06978
  [astro-ph.HE]} \BibitemShut {NoStop}%
\bibitem [{\citenamefont {Meyer}\ \emph {et~al.}(2017)\citenamefont {Meyer},
  \citenamefont {Giannotti}, \citenamefont {Mirizzi}, \citenamefont {Conrad},\
  and\ \citenamefont {S\'anchez-Conde}}]{Meyer:2016wrm}%
  \BibitemOpen
  \bibfield  {author} {\bibinfo {author} {\bibfnamefont {M.}~\bibnamefont
  {Meyer}}, \bibinfo {author} {\bibfnamefont {M.}~\bibnamefont {Giannotti}},
  \bibinfo {author} {\bibfnamefont {A.}~\bibnamefont {Mirizzi}}, \bibinfo
  {author} {\bibfnamefont {J.}~\bibnamefont {Conrad}}, \ and\ \bibinfo {author}
  {\bibfnamefont {M.~A.}\ \bibnamefont {S\'anchez-Conde}},\ }\bibfield  {title}
  {\enquote {\bibinfo {title} {{Fermi Large Area Telescope as a Galactic
  Supernovae Axionscope}},}\ }\href {\doibase 10.1103/PhysRevLett.118.011103}
  {\bibfield  {journal} {\bibinfo  {journal} {Phys. Rev. Lett.}\ }\textbf
  {\bibinfo {volume} {118}},\ \bibinfo {pages} {011103} (\bibinfo {year}
  {2017})},\ \Eprint {http://arxiv.org/abs/1609.02350} {arXiv:1609.02350
  [astro-ph.HE]} \BibitemShut {NoStop}%
\bibitem [{\citenamefont {Majumdar}\ \emph {et~al.}(2017)\citenamefont
  {Majumdar}, \citenamefont {Calore},\ and\ \citenamefont
  {Horns}}]{majumdar2017spectral}%
  \BibitemOpen
  \bibfield  {author} {\bibinfo {author} {\bibfnamefont {Jhilik}\ \bibnamefont
  {Majumdar}}, \bibinfo {author} {\bibfnamefont {Francesca}\ \bibnamefont
  {Calore}}, \ and\ \bibinfo {author} {\bibfnamefont {Dieter}\ \bibnamefont
  {Horns}},\ }\bibfield  {title} {\enquote {\bibinfo {title} {Spectral
  modulation of non-galactic plane gamma-ray pulsars due to photon-alps mixing
  in galactic magnetic field},}\ }\href@noop {} {\bibfield  {journal} {\bibinfo
   {journal} {arXiv preprint arXiv:1711.08723}\ } (\bibinfo {year}
  {2017})}\BibitemShut {NoStop}%
\bibitem [{\citenamefont {Galanti}\ and\ \citenamefont
  {Roncadelli}(2018{\natexlab{a}})}]{Galanti:2018myb}%
  \BibitemOpen
  \bibfield  {author} {\bibinfo {author} {\bibfnamefont {Giorgio}\ \bibnamefont
  {Galanti}}\ and\ \bibinfo {author} {\bibfnamefont {Marco}\ \bibnamefont
  {Roncadelli}},\ }\bibfield  {title} {\enquote {\bibinfo {title}
  {{Extragalactic photon\textendash{}axion-like particle oscillations up to
  1000 TeV}},}\ }\href {\doibase 10.1016/j.jheap.2018.07.002} {\bibfield
  {journal} {\bibinfo  {journal} {JHEAp}\ }\textbf {\bibinfo {volume} {20}},\
  \bibinfo {pages} {1--17} (\bibinfo {year} {2018}{\natexlab{a}})},\ \Eprint
  {http://arxiv.org/abs/1805.12055} {arXiv:1805.12055 [astro-ph.HE]}
  \BibitemShut {NoStop}%
\bibitem [{\citenamefont {Troitsky}(2016)}]{Troitsky:2015nxa}%
  \BibitemOpen
  \bibfield  {author} {\bibinfo {author} {\bibfnamefont {Sergey}\ \bibnamefont
  {Troitsky}},\ }\bibfield  {title} {\enquote {\bibinfo {title} {{Towards
  discrimination between galactic and intergalactic axion-photon mixing}},}\
  }\href {\doibase 10.1103/PhysRevD.93.045014} {\bibfield  {journal} {\bibinfo
  {journal} {Phys. Rev.}\ }\textbf {\bibinfo {volume} {D93}},\ \bibinfo {pages}
  {045014} (\bibinfo {year} {2016})},\ \Eprint
  {http://arxiv.org/abs/1507.08640} {arXiv:1507.08640 [astro-ph.HE]}
  \BibitemShut {NoStop}%
\bibitem [{\citenamefont {Kohri}\ and\ \citenamefont
  {Kodama}(2017)}]{Kohri:2017ljt}%
  \BibitemOpen
  \bibfield  {author} {\bibinfo {author} {\bibfnamefont {Kazunori}\
  \bibnamefont {Kohri}}\ and\ \bibinfo {author} {\bibfnamefont {Hideo}\
  \bibnamefont {Kodama}},\ }\bibfield  {title} {\enquote {\bibinfo {title}
  {{Axion-Like Particles and Recent Observations of the Cosmic Infrared
  Background Radiation}},}\ }\href {\doibase 10.1103/PhysRevD.96.051701}
  {\bibfield  {journal} {\bibinfo  {journal} {Phys. Rev.}\ }\textbf {\bibinfo
  {volume} {D96}},\ \bibinfo {pages} {051701} (\bibinfo {year} {2017})},\
  \Eprint {http://arxiv.org/abs/1704.05189} {arXiv:1704.05189 [hep-ph]}
  \BibitemShut {NoStop}%
\bibitem [{\citenamefont {Liang}\ \emph {et~al.}(2019)\citenamefont {Liang},
  \citenamefont {Zhang}, \citenamefont {Xia}, \citenamefont {Feng},
  \citenamefont {Yuan},\ and\ \citenamefont {Fan}}]{Liang:2018mqm}%
  \BibitemOpen
  \bibfield  {author} {\bibinfo {author} {\bibfnamefont {Yun-Feng}\
  \bibnamefont {Liang}}, \bibinfo {author} {\bibfnamefont {Cun}\ \bibnamefont
  {Zhang}}, \bibinfo {author} {\bibfnamefont {Zi-Qing}\ \bibnamefont {Xia}},
  \bibinfo {author} {\bibfnamefont {Lei}\ \bibnamefont {Feng}}, \bibinfo
  {author} {\bibfnamefont {Qiang}\ \bibnamefont {Yuan}}, \ and\ \bibinfo
  {author} {\bibfnamefont {Yi-Zhong}\ \bibnamefont {Fan}},\ }\bibfield  {title}
  {\enquote {\bibinfo {title} {{Constraints on axion-like particle properties
  with TeV gamma-ray observations of Galactic sources}},}\ }\href {\doibase
  10.1088/1475-7516/2019/06/042} {\bibfield  {journal} {\bibinfo  {journal}
  {JCAP}\ }\textbf {\bibinfo {volume} {1906}},\ \bibinfo {pages} {042}
  (\bibinfo {year} {2019})},\ \Eprint {http://arxiv.org/abs/1804.07186}
  {arXiv:1804.07186 [hep-ph]} \BibitemShut {NoStop}%
\bibitem [{\citenamefont {Zhang}\ \emph {et~al.}(2018)\citenamefont {Zhang},
  \citenamefont {Liang}, \citenamefont {Li}, \citenamefont {Liao},
  \citenamefont {Feng}, \citenamefont {Yuan}, \citenamefont {Fan},\ and\
  \citenamefont {Ren}}]{Zhang:2018wpc}%
  \BibitemOpen
  \bibfield  {author} {\bibinfo {author} {\bibfnamefont {Cun}\ \bibnamefont
  {Zhang}}, \bibinfo {author} {\bibfnamefont {Yun-Feng}\ \bibnamefont {Liang}},
  \bibinfo {author} {\bibfnamefont {Shang}\ \bibnamefont {Li}}, \bibinfo
  {author} {\bibfnamefont {Neng-Hui}\ \bibnamefont {Liao}}, \bibinfo {author}
  {\bibfnamefont {Lei}\ \bibnamefont {Feng}}, \bibinfo {author} {\bibfnamefont
  {Qiang}\ \bibnamefont {Yuan}}, \bibinfo {author} {\bibfnamefont {Yi-Zhong}\
  \bibnamefont {Fan}}, \ and\ \bibinfo {author} {\bibfnamefont {Zhong-Zhou}\
  \bibnamefont {Ren}},\ }\bibfield  {title} {\enquote {\bibinfo {title} {{New
  bounds on axionlike particles from the Fermi Large Area Telescope observation
  of PKS 2155-304}},}\ }\href {\doibase 10.1103/PhysRevD.97.063009} {\bibfield
  {journal} {\bibinfo  {journal} {Phys. Rev.}\ }\textbf {\bibinfo {volume}
  {D97}},\ \bibinfo {pages} {063009} (\bibinfo {year} {2018})},\ \Eprint
  {http://arxiv.org/abs/1802.08420} {arXiv:1802.08420 [hep-ph]} \BibitemShut
  {NoStop}%
\bibitem [{\citenamefont {Libanov}\ and\ \citenamefont
  {Troitsky}(2019)}]{Libanov:2019fzq}%
  \BibitemOpen
  \bibfield  {author} {\bibinfo {author} {\bibfnamefont {Maxim}\ \bibnamefont
  {Libanov}}\ and\ \bibinfo {author} {\bibfnamefont {Sergey}\ \bibnamefont
  {Troitsky}},\ }\bibfield  {title} {\enquote {\bibinfo {title} {{On the impact
  of magnetic-field models in galaxy clusters on constraints on axion-like
  paricles from the lack of irregularities in high-energy spectra of
  astrophysical sources}},}\ }\href@noop {} {\  (\bibinfo {year} {2019})},\
  \Eprint {http://arxiv.org/abs/1908.03084} {arXiv:1908.03084 [astro-ph.HE]}
  \BibitemShut {NoStop}%
\bibitem [{\citenamefont {Long}\ \emph {et~al.}(2019)\citenamefont {Long},
  \citenamefont {Lin}, \citenamefont {Tam},\ and\ \citenamefont
  {Zhu}}]{Long:2019nrz}%
  \BibitemOpen
  \bibfield  {author} {\bibinfo {author} {\bibfnamefont {G.~B.}\ \bibnamefont
  {Long}}, \bibinfo {author} {\bibfnamefont {W.~P.}\ \bibnamefont {Lin}},
  \bibinfo {author} {\bibfnamefont {P.~H.~T.}\ \bibnamefont {Tam}}, \ and\
  \bibinfo {author} {\bibfnamefont {W.~S.}\ \bibnamefont {Zhu}},\ }\bibfield
  {title} {\enquote {\bibinfo {title} {{Testing CIBER cosmic infrared
  background measurements and axionlike particles with observations of TeV
  blazars}},}\ }\href@noop {} {\  (\bibinfo {year} {2019})},\ \Eprint
  {http://arxiv.org/abs/1912.05309} {arXiv:1912.05309 [astro-ph.HE]}
  \BibitemShut {NoStop}%
\bibitem [{\citenamefont {Xia}\ \emph {et~al.}(2019)\citenamefont {Xia},
  \citenamefont {Liang}, \citenamefont {Feng}, \citenamefont {Yuan},
  \citenamefont {Fan},\ and\ \citenamefont {Wu}}]{Xia:2019yud}%
  \BibitemOpen
  \bibfield  {author} {\bibinfo {author} {\bibfnamefont {Zi-Qing}\ \bibnamefont
  {Xia}}, \bibinfo {author} {\bibfnamefont {Yun-Feng}\ \bibnamefont {Liang}},
  \bibinfo {author} {\bibfnamefont {Lei}\ \bibnamefont {Feng}}, \bibinfo
  {author} {\bibfnamefont {Qiang}\ \bibnamefont {Yuan}}, \bibinfo {author}
  {\bibfnamefont {Yi-Zhong}\ \bibnamefont {Fan}}, \ and\ \bibinfo {author}
  {\bibfnamefont {Jian}\ \bibnamefont {Wu}},\ }\bibfield  {title} {\enquote
  {\bibinfo {title} {{Searching for the possible signal of the photon-axionlike
  particle oscillation in the combined GeV and TeV spectra of supernova
  remnants}},}\ }\href {\doibase 10.1103/PhysRevD.100.123004} {\bibfield
  {journal} {\bibinfo  {journal} {Phys. Rev.}\ }\textbf {\bibinfo {volume}
  {D100}},\ \bibinfo {pages} {123004} (\bibinfo {year} {2019})},\ \Eprint
  {http://arxiv.org/abs/1911.08096} {arXiv:1911.08096 [astro-ph.HE]}
  \BibitemShut {NoStop}%
\bibitem [{\citenamefont {Rubtsov}\ and\ \citenamefont
  {Troitsky}(2014)}]{Rubtsov:2014uga}%
  \BibitemOpen
  \bibfield  {author} {\bibinfo {author} {\bibfnamefont {G.~I.}\ \bibnamefont
  {Rubtsov}}\ and\ \bibinfo {author} {\bibfnamefont {S.~V.}\ \bibnamefont
  {Troitsky}},\ }\bibfield  {title} {\enquote {\bibinfo {title} {{Breaks in
  gamma-ray spectra of distant blazars and transparency of the Universe}},}\
  }\href {\doibase 10.7868/S0370274X14180015, 10.1134/S0021364014180088}
  {\bibfield  {journal} {\bibinfo  {journal} {JETP Lett.}\ }\textbf {\bibinfo
  {volume} {100}},\ \bibinfo {pages} {355--359} (\bibinfo {year} {2014})},\
  \bibinfo {note} {[Pisma Zh. Eksp. Teor. Fiz.100,no.6,397(2014)]},\ \Eprint
  {http://arxiv.org/abs/1406.0239} {arXiv:1406.0239 [astro-ph.HE]} \BibitemShut
  {NoStop}%
\bibitem [{\citenamefont {Galanti}\ \emph {et~al.}(2019)\citenamefont
  {Galanti}, \citenamefont {Tavecchio}, \citenamefont {Roncadelli},\ and\
  \citenamefont {Evoli}}]{Galanti:2018upl}%
  \BibitemOpen
  \bibfield  {author} {\bibinfo {author} {\bibfnamefont {Giorgio}\ \bibnamefont
  {Galanti}}, \bibinfo {author} {\bibfnamefont {Fabrizio}\ \bibnamefont
  {Tavecchio}}, \bibinfo {author} {\bibfnamefont {Marco}\ \bibnamefont
  {Roncadelli}}, \ and\ \bibinfo {author} {\bibfnamefont {Carmelo}\
  \bibnamefont {Evoli}},\ }\bibfield  {title} {\enquote {\bibinfo {title}
  {{Blazar VHE spectral alterations induced by photon–ALP oscillations}},}\
  }\href {\doibase 10.1093/mnras/stz1144} {\bibfield  {journal} {\bibinfo
  {journal} {Mon. Not. Roy. Astron. Soc.}\ }\textbf {\bibinfo {volume} {487}},\
  \bibinfo {pages} {123--132} (\bibinfo {year} {2019})},\ \Eprint
  {http://arxiv.org/abs/1811.03548} {arXiv:1811.03548 [astro-ph.HE]}
  \BibitemShut {NoStop}%
\bibitem [{\citenamefont {Vogel}\ \emph {et~al.}(2017)\citenamefont {Vogel},
  \citenamefont {Laha},\ and\ \citenamefont {Meyer}}]{vogel2017diffuse}%
  \BibitemOpen
  \bibfield  {author} {\bibinfo {author} {\bibfnamefont {Hendrik}\ \bibnamefont
  {Vogel}}, \bibinfo {author} {\bibfnamefont {Ranjan}\ \bibnamefont {Laha}}, \
  and\ \bibinfo {author} {\bibfnamefont {Manuel}\ \bibnamefont {Meyer}},\
  }\bibfield  {title} {\enquote {\bibinfo {title} {Diffuse axion-like particle
  searches},}\ }\href@noop {} {\bibfield  {journal} {\bibinfo  {journal} {arXiv
  preprint arXiv:1712.01839}\ } (\bibinfo {year} {2017})}\BibitemShut {NoStop}%
\bibitem [{\citenamefont {Galanti}\ and\ \citenamefont
  {Roncadelli}(2018{\natexlab{b}})}]{Galanti:2018nvl}%
  \BibitemOpen
  \bibfield  {author} {\bibinfo {author} {\bibfnamefont {Giorgio}\ \bibnamefont
  {Galanti}}\ and\ \bibinfo {author} {\bibfnamefont {Marco}\ \bibnamefont
  {Roncadelli}},\ }\bibfield  {title} {\enquote {\bibinfo {title} {{Behavior of
  axionlike particles in smoothed out domainlike magnetic fields}},}\ }\href
  {\doibase 10.1103/PhysRevD.98.043018} {\bibfield  {journal} {\bibinfo
  {journal} {Phys. Rev. D}\ }\textbf {\bibinfo {volume} {98}},\ \bibinfo
  {pages} {043018} (\bibinfo {year} {2018}{\natexlab{b}})},\ \Eprint
  {http://arxiv.org/abs/1804.09443} {arXiv:1804.09443 [astro-ph.HE]}
  \BibitemShut {NoStop}%
\bibitem [{\citenamefont {Fayet}(1980)}]{Fayet:1980ad}%
  \BibitemOpen
  \bibfield  {author} {\bibinfo {author} {\bibfnamefont {Pierre}\ \bibnamefont
  {Fayet}},\ }\bibfield  {title} {\enquote {\bibinfo {title} {{Effects of the
  Spin 1 Partner of the Goldstino (Gravitino) on Neutral Current
  Phenomenology}},}\ }\href {\doibase 10.1016/0370-2693(80)90488-8} {\bibfield
  {journal} {\bibinfo  {journal} {Phys. Lett.}\ }\textbf {\bibinfo {volume}
  {95B}},\ \bibinfo {pages} {285--289} (\bibinfo {year} {1980})}\BibitemShut
  {NoStop}%
\bibitem [{\citenamefont {Holdom}(1986)}]{Holdom:1985ag}%
  \BibitemOpen
  \bibfield  {author} {\bibinfo {author} {\bibfnamefont {Bob}\ \bibnamefont
  {Holdom}},\ }\bibfield  {title} {\enquote {\bibinfo {title} {{Two U(1)'s and
  Epsilon Charge Shifts}},}\ }\href {\doibase 10.1016/0370-2693(86)91377-8}
  {\bibfield  {journal} {\bibinfo  {journal} {Phys. Lett.}\ }\textbf {\bibinfo
  {volume} {166B}},\ \bibinfo {pages} {196--198} (\bibinfo {year}
  {1986})}\BibitemShut {NoStop}%
\bibitem [{\citenamefont {Ruffini}\ \emph {et~al.}(2016)\citenamefont
  {Ruffini}, \citenamefont {Vereshchagin},\ and\ \citenamefont
  {Xue}}]{Ruffini:2015oha}%
  \BibitemOpen
  \bibfield  {author} {\bibinfo {author} {\bibfnamefont {R.}~\bibnamefont
  {Ruffini}}, \bibinfo {author} {\bibfnamefont {G.~V.}\ \bibnamefont
  {Vereshchagin}}, \ and\ \bibinfo {author} {\bibfnamefont {S.~S.}\
  \bibnamefont {Xue}},\ }\bibfield  {title} {\enquote {\bibinfo {title}
  {{Cosmic absorption of ultra high energy particles}},}\ }\href {\doibase
  10.1007/s10509-016-2668-5} {\bibfield  {journal} {\bibinfo  {journal}
  {Astrophys. Space Sci.}\ }\textbf {\bibinfo {volume} {361}},\ \bibinfo
  {pages} {82} (\bibinfo {year} {2016})},\ \Eprint
  {http://arxiv.org/abs/1503.07749} {arXiv:1503.07749 [astro-ph.HE]}
  \BibitemShut {NoStop}%
\bibitem [{\citenamefont {Lobanov}\ \emph {et~al.}(2013)\citenamefont
  {Lobanov}, \citenamefont {Zechlin},\ and\ \citenamefont
  {Horns}}]{Lobanov:2012pt}%
  \BibitemOpen
  \bibfield  {author} {\bibinfo {author} {\bibfnamefont {A.~P.}\ \bibnamefont
  {Lobanov}}, \bibinfo {author} {\bibfnamefont {H.~S.}\ \bibnamefont
  {Zechlin}}, \ and\ \bibinfo {author} {\bibfnamefont {D.}~\bibnamefont
  {Horns}},\ }\bibfield  {title} {\enquote {\bibinfo {title} {{Astrophysical
  searches for a hidden-photon signal in the radio regime}},}\ }\href {\doibase
  10.1103/PhysRevD.87.065004} {\bibfield  {journal} {\bibinfo  {journal} {Phys.
  Rev.}\ }\textbf {\bibinfo {volume} {D87}},\ \bibinfo {pages} {065004}
  (\bibinfo {year} {2013})},\ \Eprint {http://arxiv.org/abs/1211.6268}
  {arXiv:1211.6268 [astro-ph.CO]} \BibitemShut {NoStop}%
\bibitem [{\citenamefont {Caputo}\ \emph {et~al.}(2020)\citenamefont {Caputo},
  \citenamefont {Liu}, \citenamefont {Mishra-Sharma},\ and\ \citenamefont
  {Ruderman}}]{Caputo:2020bdy}%
  \BibitemOpen
  \bibfield  {author} {\bibinfo {author} {\bibfnamefont {Andrea}\ \bibnamefont
  {Caputo}}, \bibinfo {author} {\bibfnamefont {Hongwan}\ \bibnamefont {Liu}},
  \bibinfo {author} {\bibfnamefont {Siddharth}\ \bibnamefont {Mishra-Sharma}},
  \ and\ \bibinfo {author} {\bibfnamefont {Joshua~T.}\ \bibnamefont
  {Ruderman}},\ }\bibfield  {title} {\enquote {\bibinfo {title} {{Dark Photon
  Oscillations in Our Inhomogeneous Universe}},}\ }\href@noop {} {\  (\bibinfo
  {year} {2020})},\ \Eprint {http://arxiv.org/abs/2002.05165} {arXiv:2002.05165
  [astro-ph.CO]} \BibitemShut {NoStop}%
\bibitem [{\citenamefont {Mirizzi}\ \emph {et~al.}(2009)\citenamefont
  {Mirizzi}, \citenamefont {Redondo},\ and\ \citenamefont
  {Sigl}}]{Mirizzi:2009iz}%
  \BibitemOpen
  \bibfield  {author} {\bibinfo {author} {\bibfnamefont {Alessandro}\
  \bibnamefont {Mirizzi}}, \bibinfo {author} {\bibfnamefont {Javier}\
  \bibnamefont {Redondo}}, \ and\ \bibinfo {author} {\bibfnamefont {Gunter}\
  \bibnamefont {Sigl}},\ }\bibfield  {title} {\enquote {\bibinfo {title}
  {{Microwave Background Constraints on Mixing of Photons with Hidden
  Photons}},}\ }\href {\doibase 10.1088/1475-7516/2009/03/026} {\bibfield
  {journal} {\bibinfo  {journal} {JCAP}\ }\textbf {\bibinfo {volume} {0903}},\
  \bibinfo {pages} {026} (\bibinfo {year} {2009})},\ \Eprint
  {http://arxiv.org/abs/0901.0014} {arXiv:0901.0014 [hep-ph]} \BibitemShut
  {NoStop}%
\bibitem [{\citenamefont {Zechlin}\ \emph {et~al.}(2008)\citenamefont
  {Zechlin}, \citenamefont {Horns},\ and\ \citenamefont
  {Redondo}}]{zechlin2008new}%
  \BibitemOpen
  \bibfield  {author} {\bibinfo {author} {\bibfnamefont {Hannes-Sebastian}\
  \bibnamefont {Zechlin}}, \bibinfo {author} {\bibfnamefont {Dieter}\
  \bibnamefont {Horns}}, \ and\ \bibinfo {author} {\bibfnamefont {Javier}\
  \bibnamefont {Redondo}},\ }\bibfield  {title} {\enquote {\bibinfo {title}
  {New constraints on hidden photons using very high energy gamma-rays from the
  crab nebula},}\ }in\ \href@noop {} {\emph {\bibinfo {booktitle} {AIP
  Conference Proceedings}}},\ Vol.\ \bibinfo {volume} {1085}\ (\bibinfo
  {organization} {American Institute of Physics},\ \bibinfo {year} {2008})\
  pp.\ \bibinfo {pages} {727--730}\BibitemShut {NoStop}%
\bibitem [{\citenamefont {Kim}(1979)}]{Kim:1979if}%
  \BibitemOpen
  \bibfield  {author} {\bibinfo {author} {\bibfnamefont {Jihn~E.}\ \bibnamefont
  {Kim}},\ }\bibfield  {title} {\enquote {\bibinfo {title} {{Weak Interaction
  Singlet and Strong CP Invariance}},}\ }\href {\doibase
  10.1103/PhysRevLett.43.103} {\bibfield  {journal} {\bibinfo  {journal} {Phys.
  Rev. Lett.}\ }\textbf {\bibinfo {volume} {43}},\ \bibinfo {pages} {103}
  (\bibinfo {year} {1979})}\BibitemShut {NoStop}%
\bibitem [{\citenamefont {Shifman}\ \emph {et~al.}(1980)\citenamefont
  {Shifman}, \citenamefont {Vainshtein},\ and\ \citenamefont
  {Zakharov}}]{Shifman:1979if}%
  \BibitemOpen
  \bibfield  {author} {\bibinfo {author} {\bibfnamefont {Mikhail~A.}\
  \bibnamefont {Shifman}}, \bibinfo {author} {\bibfnamefont {A.~I.}\
  \bibnamefont {Vainshtein}}, \ and\ \bibinfo {author} {\bibfnamefont
  {Valentin~I.}\ \bibnamefont {Zakharov}},\ }\bibfield  {title} {\enquote
  {\bibinfo {title} {{Can Confinement Ensure Natural CP Invariance of Strong
  Interactions?}}}\ }\href {\doibase 10.1016/0550-3213(80)90209-6} {\bibfield
  {journal} {\bibinfo  {journal} {Nucl. Phys.}\ }\textbf {\bibinfo {volume}
  {B166}},\ \bibinfo {pages} {493--506} (\bibinfo {year} {1980})}\BibitemShut
  {NoStop}%
\bibitem [{\citenamefont {Zhitnitsky}(1980)}]{Zhitnitsky:1980tq}%
  \BibitemOpen
  \bibfield  {author} {\bibinfo {author} {\bibfnamefont {A.~R.}\ \bibnamefont
  {Zhitnitsky}},\ }\bibfield  {title} {\enquote {\bibinfo {title} {{On Possible
  Suppression of the Axion Hadron Interactions. (In Russian)}},}\ }\href@noop
  {} {\bibfield  {journal} {\bibinfo  {journal} {Sov. J. Nucl. Phys.}\ }\textbf
  {\bibinfo {volume} {31}},\ \bibinfo {pages} {260} (\bibinfo {year} {1980})},\
  \bibinfo {note} {[Yad. Fiz.31,497(1980)]}\BibitemShut {NoStop}%
\bibitem [{\citenamefont {Dine}\ \emph {et~al.}(1981)\citenamefont {Dine},
  \citenamefont {Fischler},\ and\ \citenamefont {Srednicki}}]{Dine:1981rt}%
  \BibitemOpen
  \bibfield  {author} {\bibinfo {author} {\bibfnamefont {Michael}\ \bibnamefont
  {Dine}}, \bibinfo {author} {\bibfnamefont {Willy}\ \bibnamefont {Fischler}},
  \ and\ \bibinfo {author} {\bibfnamefont {Mark}\ \bibnamefont {Srednicki}},\
  }\bibfield  {title} {\enquote {\bibinfo {title} {{A Simple Solution to the
  Strong CP Problem with a Harmless Axion}},}\ }\href {\doibase
  10.1016/0370-2693(81)90590-6} {\bibfield  {journal} {\bibinfo  {journal}
  {Phys. Lett.}\ }\textbf {\bibinfo {volume} {104B}},\ \bibinfo {pages}
  {199--202} (\bibinfo {year} {1981})}\BibitemShut {NoStop}%
\bibitem [{\citenamefont {Raffelt}\ and\ \citenamefont
  {Stodolsky}(1988{\natexlab{b}})}]{PhysRevD.37.1237}%
  \BibitemOpen
  \bibfield  {author} {\bibinfo {author} {\bibfnamefont {Georg}\ \bibnamefont
  {Raffelt}}\ and\ \bibinfo {author} {\bibfnamefont {Leo}\ \bibnamefont
  {Stodolsky}},\ }\bibfield  {title} {\enquote {\bibinfo {title} {Mixing of the
  photon with low-mass particles},}\ }\href {\doibase 10.1103/PhysRevD.37.1237}
  {\bibfield  {journal} {\bibinfo  {journal} {Phys. Rev. D}\ }\textbf {\bibinfo
  {volume} {37}},\ \bibinfo {pages} {1237--1249} (\bibinfo {year}
  {1988}{\natexlab{b}})}\BibitemShut {NoStop}%
\bibitem [{\citenamefont {Meyer}\ \emph {et~al.}(2014)\citenamefont {Meyer},
  \citenamefont {Montanino},\ and\ \citenamefont {Conrad}}]{Meyer:2014epa}%
  \BibitemOpen
  \bibfield  {author} {\bibinfo {author} {\bibfnamefont {Manuel}\ \bibnamefont
  {Meyer}}, \bibinfo {author} {\bibfnamefont {Daniele}\ \bibnamefont
  {Montanino}}, \ and\ \bibinfo {author} {\bibfnamefont {Jan}\ \bibnamefont
  {Conrad}},\ }\bibfield  {title} {\enquote {\bibinfo {title} {{On detecting
  oscillations of gamma rays into axion-like particles in turbulent and
  coherent magnetic fields}},}\ }\href {\doibase 10.1088/1475-7516/2014/09/003}
  {\bibfield  {journal} {\bibinfo  {journal} {JCAP}\ }\textbf {\bibinfo
  {volume} {09}},\ \bibinfo {pages} {003} (\bibinfo {year} {2014})},\ \Eprint
  {http://arxiv.org/abs/1406.5972} {arXiv:1406.5972 [astro-ph.HE]} \BibitemShut
  {NoStop}%
\bibitem [{\citenamefont {Dobrynina}\ \emph {et~al.}(2015)\citenamefont
  {Dobrynina}, \citenamefont {Kartavtsev},\ and\ \citenamefont
  {Raffelt}}]{Dobrynina:2014qba}%
  \BibitemOpen
  \bibfield  {author} {\bibinfo {author} {\bibfnamefont {Alexandra}\
  \bibnamefont {Dobrynina}}, \bibinfo {author} {\bibfnamefont {Alexander}\
  \bibnamefont {Kartavtsev}}, \ and\ \bibinfo {author} {\bibfnamefont {Georg}\
  \bibnamefont {Raffelt}},\ }\bibfield  {title} {\enquote {\bibinfo {title}
  {{Photon-photon dispersion of TeV gamma rays and its role for photon-ALP
  conversion}},}\ }\href {\doibase 10.1103/PhysRevD.91.083003} {\bibfield
  {journal} {\bibinfo  {journal} {Phys. Rev. D}\ }\textbf {\bibinfo {volume}
  {91}},\ \bibinfo {pages} {083003} (\bibinfo {year} {2015})},\ \bibinfo {note}
  {[Erratum: Phys.Rev.D 95, 109905 (2017)]},\ \Eprint
  {http://arxiv.org/abs/1412.4777} {arXiv:1412.4777 [astro-ph.HE]} \BibitemShut
  {NoStop}%
\bibitem [{\citenamefont {Vernetto}\ and\ \citenamefont
  {Lipari}(2016)}]{Vernetto:2016alq}%
  \BibitemOpen
  \bibfield  {author} {\bibinfo {author} {\bibfnamefont {Silvia}\ \bibnamefont
  {Vernetto}}\ and\ \bibinfo {author} {\bibfnamefont {Paolo}\ \bibnamefont
  {Lipari}},\ }\bibfield  {title} {\enquote {\bibinfo {title} {{Absorption of
  very high energy gamma rays in the Milky Way}},}\ }\href {\doibase
  10.1103/PhysRevD.94.063009} {\bibfield  {journal} {\bibinfo  {journal} {Phys.
  Rev.}\ }\textbf {\bibinfo {volume} {D94}},\ \bibinfo {pages} {063009}
  (\bibinfo {year} {2016})},\ \Eprint {http://arxiv.org/abs/1608.01587}
  {arXiv:1608.01587 [astro-ph.HE]} \BibitemShut {NoStop}%
\bibitem [{\citenamefont {Jansson}\ and\ \citenamefont
  {Farrar}(2012)}]{Jansson:2012rt}%
  \BibitemOpen
  \bibfield  {author} {\bibinfo {author} {\bibfnamefont {Ronnie}\ \bibnamefont
  {Jansson}}\ and\ \bibinfo {author} {\bibfnamefont {Glennys~R.}\ \bibnamefont
  {Farrar}},\ }\bibfield  {title} {\enquote {\bibinfo {title} {{The Galactic
  Magnetic Field}},}\ }\href {\doibase 10.1088/2041-8205/761/1/L11} {\bibfield
  {journal} {\bibinfo  {journal} {Astrophys. J.}\ }\textbf {\bibinfo {volume}
  {761}},\ \bibinfo {pages} {L11} (\bibinfo {year} {2012})},\ \Eprint
  {http://arxiv.org/abs/1210.7820} {arXiv:1210.7820 [astro-ph.GA]} \BibitemShut
  {NoStop}%
\bibitem [{\citenamefont {Fortin}\ and\ \citenamefont
  {Sinha}(2019)}]{Fortin:2019npr}%
  \BibitemOpen
  \bibfield  {author} {\bibinfo {author} {\bibfnamefont {Jean-François}\
  \bibnamefont {Fortin}}\ and\ \bibinfo {author} {\bibfnamefont {Kuver}\
  \bibnamefont {Sinha}},\ }\bibfield  {title} {\enquote {\bibinfo {title}
  {{Photon-Dark Photon Conversions in Background Electromagnetic Fields}},}\
  }\href@noop {} {\  (\bibinfo {year} {2019})},\ \Eprint
  {http://arxiv.org/abs/1904.08968} {arXiv:1904.08968 [hep-ph]} \BibitemShut
  {NoStop}%
\bibitem [{\citenamefont {Zaborov}\ \emph {et~al.}(2017)\citenamefont
  {Zaborov}, \citenamefont {Taylor}, \citenamefont {Sanchez}, \citenamefont
  {Lenain},\ and\ \citenamefont {Romoli}}]{Zaborov:2016jub}%
  \BibitemOpen
  \bibfield  {author} {\bibinfo {author} {\bibfnamefont {D.}~\bibnamefont
  {Zaborov}}, \bibinfo {author} {\bibfnamefont {A.~M.}\ \bibnamefont {Taylor}},
  \bibinfo {author} {\bibfnamefont {D.~A.}\ \bibnamefont {Sanchez}}, \bibinfo
  {author} {\bibfnamefont {J.~P.}\ \bibnamefont {Lenain}}, \ and\ \bibinfo
  {author} {\bibfnamefont {C.}~\bibnamefont {Romoli}} (\bibinfo {collaboration}
  {H.E.S.S.}),\ }\bibfield  {title} {\enquote {\bibinfo {title} {{Gamma-ray
  blazar spectra with H.E.S.S. II mono analysis: The case of PKS
  2155\ensuremath{-}304 and PG 1553+113}},}\ }\href {\doibase
  10.1063/1.4968963} {\bibfield  {journal} {\bibinfo  {journal} {AIP Conf.
  Proc.}\ }\textbf {\bibinfo {volume} {1792}},\ \bibinfo {pages} {050017}
  (\bibinfo {year} {2017})},\ \Eprint {http://arxiv.org/abs/1612.05111}
  {arXiv:1612.05111 [astro-ph.HE]} \BibitemShut {NoStop}%
\bibitem [{\citenamefont {Anastassopoulos}\ \emph {et~al.}(2017)\citenamefont
  {Anastassopoulos} \emph {et~al.}}]{Anastassopoulos:2017ftl}%
  \BibitemOpen
  \bibfield  {author} {\bibinfo {author} {\bibfnamefont {V.}~\bibnamefont
  {Anastassopoulos}} \emph {et~al.} (\bibinfo {collaboration} {CAST}),\
  }\bibfield  {title} {\enquote {\bibinfo {title} {{New CAST Limit on the
  Axion-Photon Interaction}},}\ }\href {\doibase 10.1038/nphys4109} {\bibfield
  {journal} {\bibinfo  {journal} {Nature Phys.}\ }\textbf {\bibinfo {volume}
  {13}},\ \bibinfo {pages} {584--590} (\bibinfo {year} {2017})},\ \Eprint
  {http://arxiv.org/abs/1705.02290} {arXiv:1705.02290 [hep-ex]} \BibitemShut
  {NoStop}%
\bibitem [{\citenamefont {Essig}\ \emph {et~al.}(2013)\citenamefont {Essig},
  \citenamefont {Jaros}, \citenamefont {Wester}, \citenamefont {Adrian},
  \citenamefont {Andreas}, \citenamefont {Averett}, \citenamefont {Baker},
  \citenamefont {Batell}, \citenamefont {Battaglieri}, \citenamefont {Beacham}
  \emph {et~al.}}]{essig2013dark}%
  \BibitemOpen
  \bibfield  {author} {\bibinfo {author} {\bibfnamefont {Rouven}\ \bibnamefont
  {Essig}}, \bibinfo {author} {\bibfnamefont {John~A}\ \bibnamefont {Jaros}},
  \bibinfo {author} {\bibfnamefont {William}\ \bibnamefont {Wester}}, \bibinfo
  {author} {\bibfnamefont {P~Hansson}\ \bibnamefont {Adrian}}, \bibinfo
  {author} {\bibfnamefont {S}~\bibnamefont {Andreas}}, \bibinfo {author}
  {\bibfnamefont {T}~\bibnamefont {Averett}}, \bibinfo {author} {\bibfnamefont
  {O}~\bibnamefont {Baker}}, \bibinfo {author} {\bibfnamefont {B}~\bibnamefont
  {Batell}}, \bibinfo {author} {\bibfnamefont {M}~\bibnamefont {Battaglieri}},
  \bibinfo {author} {\bibfnamefont {J}~\bibnamefont {Beacham}},  \emph
  {et~al.},\ }\bibfield  {title} {\enquote {\bibinfo {title} {Dark sectors and
  new, light, weakly-coupled particles},}\ }\href@noop {} {\bibfield  {journal}
  {\bibinfo  {journal} {arXiv preprint arXiv:1311.0029}\ } (\bibinfo {year}
  {2013})}\BibitemShut {NoStop}%
\end{thebibliography}%

\appendix

\section{Flux binning \& scaling}
\label{app:binning}

The energy uncertainty of terrestrial cosmic ray experiments is generally non-negligible thus the smearing of the observed energy needs to be incorporated. 
The energy resolution of AS$\gamma$ ranges from 20\% to 40\%\cite{Amenomori:2019rjd} and we adopt 30\% for this analysis. 
Similarly, the energy resolution for HEGRA, MAGIC and HAWC is taken as $\delta=$10\% \cite{Aharonian:2004gb}, 16\%~\cite{Aleksic:2014lkm} and 23\%~\cite{Abeysekara:2019edl} of the detected energy.
The observed spectrum is then convoluted with a normally distributed energy around the `true' energy from the incoming spectrum.

For a detector measuring a photon in one energy bin ranging from $E_{0}$ to $E_{0}+\Delta E$, the expected flux is
\be
\Delta \phi = \int^{E_0+\Delta E}_{E_0} {\rm d}E\int^{\infty}_{0} A(E^\prime, E)\frac{{\rm d} \phi}{{\rm d}E^\prime}{\rm d}E^\prime
\ee
where $E$ and $E'$ are the observed and `true' cosmic energies, $A$ is a window function that takes account of $E'$ being observed at $E$ with a normally distributed probability with energy uncertainty $\delta\cdot E$, and $\frac{{\rm d} \phi}{{\rm d}E^\prime}$ is the incoming differential energy flux. 
After integrating over the energy bin the expected flux is,
\begin{eqnarray}
\Delta \phi=\int^{\infty}_{0} \tilde{A}(E^\prime, E_0, \Delta E)\frac{{\rm d} \phi}{{\rm d}E^\prime}dE^\prime,
\end{eqnarray}
where the integrated $\tilde{A}(E^\prime$, $E_0$, $E_0+\Delta E)$ takes the form
\begin{equation}
    \tilde{A}=\frac{1}{2}\left[{\rm erf}\left(\frac{E_0+\Delta E-E^\prime}{\sqrt{2}\delta\cdot E^\prime}\right)-{\rm erf}\left(\frac{E_0-E^\prime}{\sqrt{2}\delta\cdot E^\prime}\right)\right],
\end{equation}
where `erf' is the usual Gaussian error function. After spectral smearing, we still need to consider an overall spectral shift due to the experimental energy scaling uncertainty, $E\rightarrow f\cdot E$. The flux in an $f-$shifted energy bin is
\begin{eqnarray}
\Delta \phi=\int^{\infty}_{0} \tilde{A}(E^\prime, f\cdot E_0, f\cdot \Delta E)f^n\cdot\frac{{\rm d} \phi}{{\rm d}E^\prime}dE^\prime.
\end{eqnarray}

Often the experimental data are given in the form of binned flux multiplied by $E^n$. Without assumptions on the incoming spectrum, shifted experimental data are treated as $E^n\frac{{\rm d}\phi}{{\rm d}E}\rightarrow f^{n-1}\cdot E^n\frac{{\rm d}\phi}{{\rm d}E}$, maintaining the same event counts in the (shifted) energy bin.

\begin{figure}
\includegraphics[scale=0.85]{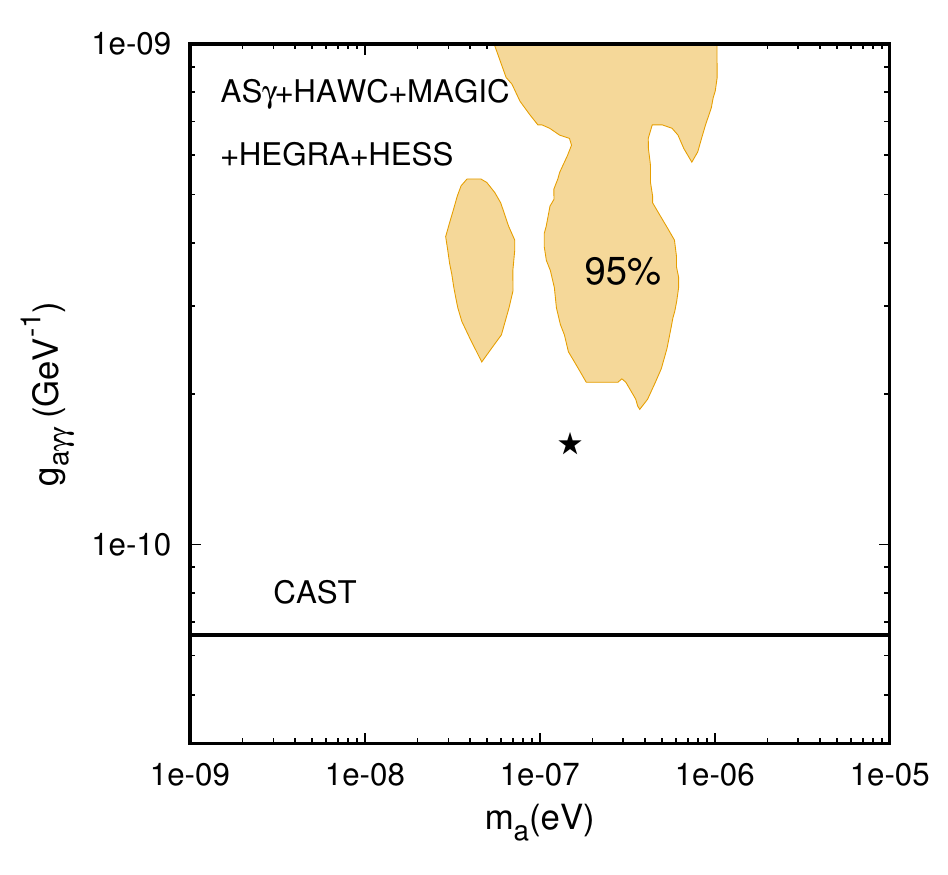}
\caption{Similar to Fig.~\ref{fig:fit_axion}, ALP oscillation limits with all five data sets.
The best fit $\chi^2_{\rm min}$=57.8 is marked by the asterisk point.}
\label{fig:5set_axion}
\end{figure}

\section{Comparisons with HESS data}
\label{app:comparisons}

Here we list the IC background fitting result to individual experiments, and also the ALP-oscillated fitting results after including HESS data into the joint analysis. 

\begin{table}[h]
\caption{Inverse-Compton spectrum fits to each experiment. $\phi_0$ and $E_0$ take unit of TeV cm$^2$s$^{-1}$ and TeV.}
\begin{tabular}{c|c|c|cccc}
\hline\hline
 Data & $\chi^2_{\rm min}$ & d.o.f. & $\phi_0$ & $E_0$ & $\alpha$ & $\beta$\\
\hline
 Tibet AS$\gamma$ ~\cite{Amenomori:2019rjd} &1.8 &6 &9.28$\times 10^{-12}$&1.47 &-2.12 &-0.32 \\
 HEGRA~\cite{Aharonian:2004gb} &13.6&12 &1.58 $\times 10^{-12}$&3.02 &-2.65  &-0.11 \\
 MAGIC~\cite{Aleksic:2014lkm} &6.6 &7 &9.96$\times 10^{-13}$ &3.79 &-2.80 &-0.25 \\
 HESS~\cite{Abramowski:2013qea} &14.8 &28 &2.04$\times 10^{-13}$  &6.44 &-2.79&-4.06$\times 10^{-3}$ \\
 HAWC~\cite{Abeysekara:2019edl} &6.5 &5 &1.68$\times 10^{-11}$  &1.44 &-2.49&-0.16 \\
\hline\hline
\end{tabular}
\label{tab:each_set}
\end{table}

Table.~\ref{tab:each_set} shows the dataset background consistency with the best fitting results to Inverse Compton spectrum in Eq.~(\ref{eq:parabola})  after considering the absorption effect due to the background photons. 
The reduced $\chi^2$ values from AS$\gamma$ and HESS fits are significantly less than one, indicating their combination with other data sets may loosen constraints on spectral distortions. 
As the new AS$\gamma$ data above 100 TeV are necessary for testing axions at higher mass ranges, we only include AS$\gamma$ in Section~\ref{sect:analysis}. The five-set joint result with HESS included is shown in Fig.~\ref{fig:5set_axion}, where the 95\% exclusion region moves slightly to larger $g_{a\gamma\gamma}$.

\end{document}